\documentclass[twocolumn]{aastex631}

\newcommand{\lapp}{$_<\atop{^\sim}$}

\newcommand{\msun}{$M_{\odot}$}
\newcommand{\lsun}{$L_{\odot}$}
\newcommand{\h}{$h^{-1}$}
\newcommand{\kms}{km~s$^{-1}$}

\newcommand{\hi}{H{\sc\,i}}

\shortauthors{Rhode et al.}
\shorttitle{Structural Properties of Andromeda Satellites Cas III, Per I, \& Lac I}

\begin{document}

\title{Exploring the Structures and Substructures of the Andromeda Satellite Dwarf Galaxies \\ Cassiopeia~III, Perseus~I, and Lacerta~I}

\author[0000-0001-8283-4591]{Katherine L. Rhode}
\affiliation{Department of Astronomy, Indiana University, 727 East
 Third Street, Bloomington, IN 47405, USA} 
\email{krhode@indiana.edu}

\author[0000-0002-3222-2949]{Nicholas J. Smith}\affiliation{Department of Astronomy, Indiana University, 727 East
 Third Street, Bloomington, IN 47405, USA} 

\author[0000-0002-1763-4128]{Denija Crnojevic}
\affiliation{University of Tampa, 401 West Kennedy Boulevard, Tampa, FL 33606, USA} 

\author[0000-0003-4102-380X]{David J. Sand}
\affiliation{Steward Observatory, University of Arizona, 933 North Cherry Avenue, Room N204, Tucson, AZ 85721, USA} 

\author[0000-0003-1548-4602]{Ryan A. Lambert}
\affiliation{Department of Astronomy, Indiana University, 727 East
 Third Street, Bloomington, IN 47405, USA} 

\author[0000-0003-2742-6872]{Enrico Vesperini}
\affiliation{Department of Astronomy, Indiana University, 727 East
 Third Street, Bloomington, IN 47405, USA} 

\author[0000-0002-1813-4053]{Madison V. Smith}
\affiliation{Department of Astronomy, Indiana University, 727 East
 Third Street, Bloomington, IN 47405, USA} 

\author[0000-0001-9165-8905]{Steven Janowiecki}
\affiliation{University of Texas, Hobby-Eberly Telescope, McDonald Observatory, TX 79734, USA}

\author[0000-0001-8483-603X]{John J. Salzer}
\affiliation{Department of Astronomy, Indiana University, 727 East
 Third Street, Bloomington, IN 47405, USA}

\author[0000-0001-8855-3635]{Ananthan Karunakaran}
\affiliation{Instituto de Astrofisica  de Andalucia (CSIC), Glorieta de la Astronomia, E-18008 Granada, Spain}

\author[0000-0002-0956-7949]{Kristine Spekkens}
\affiliation{Department of Physics and Space Science, Royal Military College of Canada, P.O. Box 17000, Station Forces Kingston, ON K7K 7B4, Canada}
\affiliation{Department of Physics, Engineering Physics and Astronomy, Queen’s University, Kingston, ON K7L 3N6, Canada}
 
\begin{abstract}
We present results from wide-field imaging of the resolved stellar
populations of the dwarf spheroidal galaxies Cassiopeia~III (And~XXXII) and
Perseus~I (And~XXXIII), two satellites in the outer stellar halo of
the Andromeda galaxy (M31).  Our WIYN pODI photometry traces the red
giant star population in each galaxy to $\sim$2.5$-$3 half-light radii
from the galaxy center.  We use the Tip of the Red Giant Branch (TRGB)
method to derive distances of $(m-M)_0$ $=$ 24.62$\pm$0.12 mag
(839$^{+48}_{-45}$ kpc, or 156$^{+16}_{-13}$ kpc 
from M31) for Cas~III and 24.47$\pm$0.13 mag (738$^{+48}_{-45}$ kpc, or 
351$^{+17}_{-16}$ kpc 
from M31) for Per~I. These values are consistent within the errors
with TRGB distances derived from a deeper Hubble Space Telescope study
of the galaxies' inner regions. For each galaxy, we derive structural
parameters, total magnitude, and central surface brightness.  We also
place upper limits on the ratio of neutral hydrogen gas mass to
optical luminosity, confirming the gas-poor nature of both galaxies.
We combine our data set with corresponding data for the M31 satellite
galaxy Lacerta~I (And~XXXI) from earlier work, and search for
substructure within the RGB star populations of Cas~III,
Per~I, and Lac~I. We find an overdense region on the west side of
Lac~I at a significance level of 2.5--3-$\sigma$ and a
low-significance filament extending in the direction of M31.  In
Cas~III, we identify two modestly significant overdensities near the
center of the galaxy and another at two half-light radii.  Per~I shows no
evidence for substructure in its RGB star population, which may
reflect this galaxy's isolated nature.
\end{abstract}


\section{Introduction}
\label{sec: introduction}

Systematic surveys of our local galaxy neighborhood carried out over the past few decades have led to the discovery of dozens of low-mass galaxies in  the Local Group and its immediate environs \citep[e.g.,][and references therein]{mcconnachie12,simon19}. Follow-up 
studies of these newly-identified dwarf galaxies have in turn yielded valuable insights into fundamental astrophysical topics such as the evolution of the stellar and gaseous components of low-mass systems \citep[e.g.,][]{weisz14,mcquinn15a,putman21}, 
the interplay between dwarf galaxies and their surrounding environments \citep[e.g.,][]{mcquinn15b,fillingham18}, the dynamics and dark matter content of low-mass galaxies
\citep[e.g.,][]{simon07,simon11,kirby14,wheeler15,brownsberger21}, and the formation and evolution of structure over cosmic history in regions like the Local Group  \citep[e.g.,][]{garrison-kimmel14,verbeke15,revaz18,applebaum21}. 
Over the past two decades, special attention has been paid to searching for faint dwarf galaxy satellites around our nearest massive galaxy neighbor, Andromeda (M31). For example, searches of Sloan Digital Sky Survey (SDSS) imaging of certain regions around M31 have led to the discovery of several faint dwarf satellites
with absolute $V$-band magnitudes $M_V$ of $\sim$ $-$8 \citep[e.g.,][]{zucker04,zucker07,slater11,bell11}.
An imaging study by \citet{irwin08} used data from a wide-field camera on the Isaac Newton Telescope to find a 
dwarf spheroidal galaxy with $M_V$ $\sim$ $-$8.5, located only $\sim$40~kpc in projection from M31.  The Pan-Andromeda Archaeological Survey \citep[PAndAS;][]{mcconnachie09}, 
a dedicated survey carried out with the 
Canada-France-Hawaii Telescope (CFHT)
that imaged the area within 
a projected distance of $\sim$150~kpc around Andromeda, 
has revealed more than a dozen dwarf satellite galaxies with a range of luminosities ($M_V$ ranging from $\sim$ $-$6 to $-$10)
and projected distances from M31
\citep{martin06,ibata07,mcconnachie08,martin09,richardson11,mcconnachie18}. 

The Pan-STARRS 3$\pi$ survey, which imaged the sky at declinations north of $-$30$\degr$ in five optical to near-infrared bands, enabled the discovery of even more Andromeda satellite dwarf galaxies.
Martin and collaborators used the photometric catalogs from the 3$\pi$ survey to uncover three previously unknown dwarf spheroidal galaxies located in regions that had either never been systematically surveyed, or had only shallow coverage:
Lacerta I (And XXXI), Cassiopeia III (And XXXII), and Perseus I (And XXXIII) (\citealt{martin13a}, \citealt{martin13b}). The three galaxies were estimated to have total absolute $V$-band magnitudes $M_V$ of $-$11.7, $-$12.3, and $-$10.3, respectively \citep{martin13a,martin13b}, placing them squarely in the mid-range for Local Group dwarf satellite galaxies \citep[e.g.,][]{mcconnachie12}.  However, all three have 
relatively faint central surface brightnesses ($\mu_V$ $\sim$ 26~mag~arcsec$^{-2}$) and 
are more than 10~degrees away from M31 on the sky, 
qualities which help explain why they remained undiscovered until recently.

\cite{martin14} used spectroscopic follow-up measurements of the stars in all three galaxies to confirm that all 
are satellites of M31 and to measure their average stellar metallicity, finding [Fe/H] values of $-$2.0$\pm$0.1, 
$-$1.7$\pm$0.1, and $-$2.0$\pm$0.2 for Lac~I, Cas~III, and Per~I respectively.  \citet{martin14} also measured velocity dispersions and dynamical masses for the two more massive galaxies, Lac~I and Cas~III, and concluded that these properties are typical of 
those 
of other Andromeda dwarf satellite galaxies with similar luminosities and sizes. 
A feature that distinguishes Lac~I, Cas~III, and Per~I from other Andromeda satellites is their 
large distances from the center of their parent galaxy
\citep[between $\sim$150 and $\sim$350 kpc;][]{martin13a,martin13b,weisz19a,savino22a}
which makes them particularly useful 
tracers of the host galaxy's 
mass distribution \citep{martin14, watkins10}. \citet{martin14} also note that while Cas~III has a location that puts it within the thin plane of satellite galaxies co-rotating with Andromeda that was identified by \citet{ibata13}, its velocity indicates it is moving in the opposite direction compared to the bulk motion of galaxies in that structure.

The 
photometry from Pan-STARRS that \citet{martin13a} and \citet{martin13b} used to discover Lac~I, Cas~III, and Per~I covered a wide field but was fairly shallow, reaching to approximately $i_{Pi}$ $\sim$ 21$-$22.  Our group followed up with a deeper ($i$ $\sim$24$-$25) wide-field photometric study of Lac~I, carried out with the WIYN 3.5-m telescope\footnote{The WIYN Observatory is a joint facility of the NSF’s National Optical-Infrared Astronomy Research Laboratory, Indiana University,
the University of Wisconsin-Madison, Pennsylvania State University,
and
Purdue University.} and the 24$\arcmin$ x 24$\arcmin$ pODI camera \citep{rhode17}. We used the WIYN imaging data to 
trace 
Lac~I's stellar distribution 
beyond two times the half-light radius $r_h$, estimate the galaxy's structural parameters, derive a more precise Tip of the Red Giant Branch (TRGB) distance,
and investigate the 
stellar metallicity distribution at different radii in the galaxy. 

\cite{martin17a} utilized the Hubble Space Telescope Advanced Camera for Surveys (HST ACS) to carry out substantially deeper photometry in the central few arc~minutes of Lac~I, Cas~III, Per~I and 14 other M31 dwarf satellite galaxies, producing high-quality color-magnitude diagrams that reach well below the Horizontal Branch (HB). Their HST data allowed them to characterize the galaxies' HB morphologies and investigate their star formation histories (SFHs); they conclude that the red HBs seen in most of the sample suggest that the M31 satellites had more extended SFHs than their counterparts in the Milky Way.  In a follow-up paper in the same series, \citet{weisz19a} used the HST data to derive both TRGB and HB distances to 
Lac~I, Cas~III, and Per~I and the other galaxies in their sample.  
The HST values are, overall, in good agreement with the earlier ground-based distance estimates but are much more precise 
(see Section~\ref{sec: trgb} of the current paper for a more detailed comparison of  the various distance measurements).  \citet{weisz19b} then examined the star formation quenching times of the galaxies in their HST data set along with a few others for which archival HST observations exist.  They found important differences between the M31 satellites' formation histories and those of the Milky Way's satellites -- the M31 dwarfs they studied (including Lac~I, Cas~III and Per~I) quenched their star formation $\sim$3$-$9~Gyr ago, instead of showing a more bimodal distribution of very early ($\sim$10$-$12~Gyr ago) or very late (\lapp 3~Gyr ago) quenching times like the Milky Way dwarf population. They point out that these differences may reflect the different accretion histories of the Galaxy and Andromeda. 

Two other recent studies 
are relevant to mention here.
Per~I was observed by \citet{higgs21}, who included it in the Solitary Local dwarfs (Solo) survey \citep{higgs16}, a volume-limited survey of dwarf galaxies within 3~Mpc that are located more than 300~kpc from the Milky Way or M31.  \citet{higgs21} used wide-field $g$ and $i$ imaging and photometry from the CFHT and Magellan telescopes to quantify the distances, spatial distributions, and morphologies of Per~I and several other 
isolated dwarfs in the Local Group. They generally found agreement between their work and previous measurements, and found no strong evidence of extended stellar substructures around Per~I or the other 11 galaxies in their sample. 
Most recently, \citet{savino22a} utilized deep imaging obtained as part of the HST Treasury Survey of the M31 satellite galaxies 
to identify RR Lyrae variable stars and derive accurate distances for M31 and 38 of its associated stellar systems, including Lac~I, Cas~III, and Per~I.  They use their  measurements to map out the 3D structure of the M31 satellite system.  In the process, they compare their distance determinations to TRGB distances in the literature and find reasonable consistency, although they point out that especially for faint galaxies (with $M_V$ $>$ $-$9.5), the sparsely populated nature of some portions of the color-magnitude diagram (CMD) can lead to distances that are systematically overestimated compared to the more accurate distances derived from RR~Lyrae stars.

The current paper is a follow-up to the WIYN imaging study of the photometric and structural properties of Lac~I that was presented in \citet{rhode17}.  Here we present results from WIYN imaging of Cas~III and Per~I that reaches a similar depth ($i$ $\sim$24$-$25) and covers a $\sim$20\arcmin\ x 20\arcmin\ area around each galaxy.  Although our photometry is much shallower than the photometry from the various HST studies, the wide field allows us to trace the stellar populations out to $\sim$10\arcmin\ ($\sim$2$-$3 half-light radii).  We use the photometry to construct a CMD for the galaxies and then combine our Cas~III and Per~I data with the data for Lac~I from \citet{rhode17} to carry out a detailed examination of the structures and substructures within and around all three stellar systems.

The paper is organized as follows.  In Sections~\ref{sec: observations} and \ref{sec: photometry}, we describe the observations, reductions, and initial analysis (source detection and photometry) 
carried out with the WIYN data.  In Section~\ref{sec: properties}, we present the CMD for the stars in each galaxy and use it to characterize the stellar populations and derive a TRGB distance, which we then compare to the other distance measurements in the literature.  We also use our data to derive structural parameters, total magnitude, and central surface brightness for Cas~III and Per~I, and we combine our optical measurements with published estimates of the galaxies' HI masses to constrain the \hi\ mass to optical luminosity ratio of the galaxies. 
In Section~\ref{sec: substructure} we present the results from a search for substructure that we carried out using the spatial distributions of the stars in all three galaxies in the WIYN data set (Lac~I, Cas~III, and Per~I).  
The last section of the paper summarizes our main conclusions. 

\section{Observations and Data Reduction}
\label{sec: observations}
Observations of Cas~III and Per~I were obtained on 2-3 October 2013 and 29-31 October 2013, respectively, with the WIYN 3.5-m telescope and the One Degree Imager camera with a partially filled focal plane (pODI; \citealt{harbeck14}). The pODI instrument included a central 3 x 3 array of Orthogonal Transfer Arrays (OTAs) that was used as the main imaging area, and a few outlying OTAs used for guiding.  Each of the OTAs was made up of an 8 x 8 array of orthogonal transfer CCDs.  The field-of-view of the central imaging area 
in pODI was $\sim$24\arcmin\ x 24\arcmin, with a pixel scale of 0.11\arcsec\ per pixel.  

The galaxies were observed in the SDSS $g$ and $i$ filters.  
Several 600$-$700~s exposures were taken of each galaxy in each filter over the course of the observing run. The telescope was dithered between exposures in order to fill in the gaps between the CCDs and OTAs in the final science images.  The individual images were assessed and the frames 
with the best image quality were chosen for inclusion in the stacked images of a given galaxy.  
For Cas~III, the total exposure time 
in the stacked images was 9000~s in $g$ (13 exposures) and 5000~s in $i$ (8 exposures); for Per~I, the total exposure time was 5400~s in $g$ (9 exposures) and 8400~s in $i$ (14 exposures).

All of the pODI images were ingested into the ODI Pipeline, Portal, and Archive (ODI-PPA, \citealt{gopu14})\footnote{The ODI Pipeline, Portal, and Archive (ODI-PPA) system is a joint development project of the WIYN Consortium, Inc., in partnership with Indiana University's Pervasive Technology Institute (PTI) and NSF’s NOIRLab.}
and then processed with the QuickReduce pipeline \citep{kotulla14} to mask saturated pixels, apply crosstalk and persistence corrections, subtract overscan signal, correct for non-linearity, apply bias, dark, and flatfield, and pupil ghost corrections, and remove cosmic rays. After the QuickReduce step, the images were illumination-corrected, reprojected to the same pixel scale, and scaled to a common flux level before being combined into a single, deep science-ready image in each filter.  The edges of the images where the noise level was higher (due to incomplete coverage from the dither pattern) were trimmed, yielding a final image size of $\sim$20\arcmin\ $\times$ 20\arcmin. For the stacked images of Cas~III, the mean Full Width at Half Maximum of the point-spread function (FWHM PSF) is 0.8\arcsec\ in both $g$ and $i$.  
For the combined Per~I images, the average FWHM PSF value was 0.8\arcsec\ in \textit{g} and 0.9\arcsec\ in \textit{i}.

Photometric calibration coefficients (zero points and color terms) were calculated 
using measurements of SDSS stars that appeared within the final combined pODI images of Per~I.  
Because Cas~III lies outside of the SDSS footprint, the pODI images of the galaxy were instead calibrated using traditional methods. Specifically, we took modest-length exposures of Cas~III as well as a series of standard star fields on a night during the observing run that had 
photometric sky conditions.  We used the standard star fields to calculate photometric coefficients that were used to calibrate the galaxy images taken on the photometric night.  We then calculated zero point offsets between the final combined Cas~III images and the 
calibrated images from the photometric night.
The appropriate photometric coefficients were later applied to our measurements of other sources in the final science images of Per~I and Cas~III, to yield photometrically-calibrated magnitudes and colors.  
We also calculated individual extinction corrections for each source based on its position in the images by using the \citet{schlafly11} coefficients with the \citet{schlegel98} maps of Galactic dust emission, and applied those values to yield dereddened $g,i$ magnitudes and colors. The mean color excess for the Cas~III field is $E(B-V)$ $\sim$0.2 and for the Per~I field it is $E(B-V)$ $\sim$0.15. 

\section{Source Detection and PSF Photometry}
\label{sec: photometry}

\begin{figure*}
\plotone{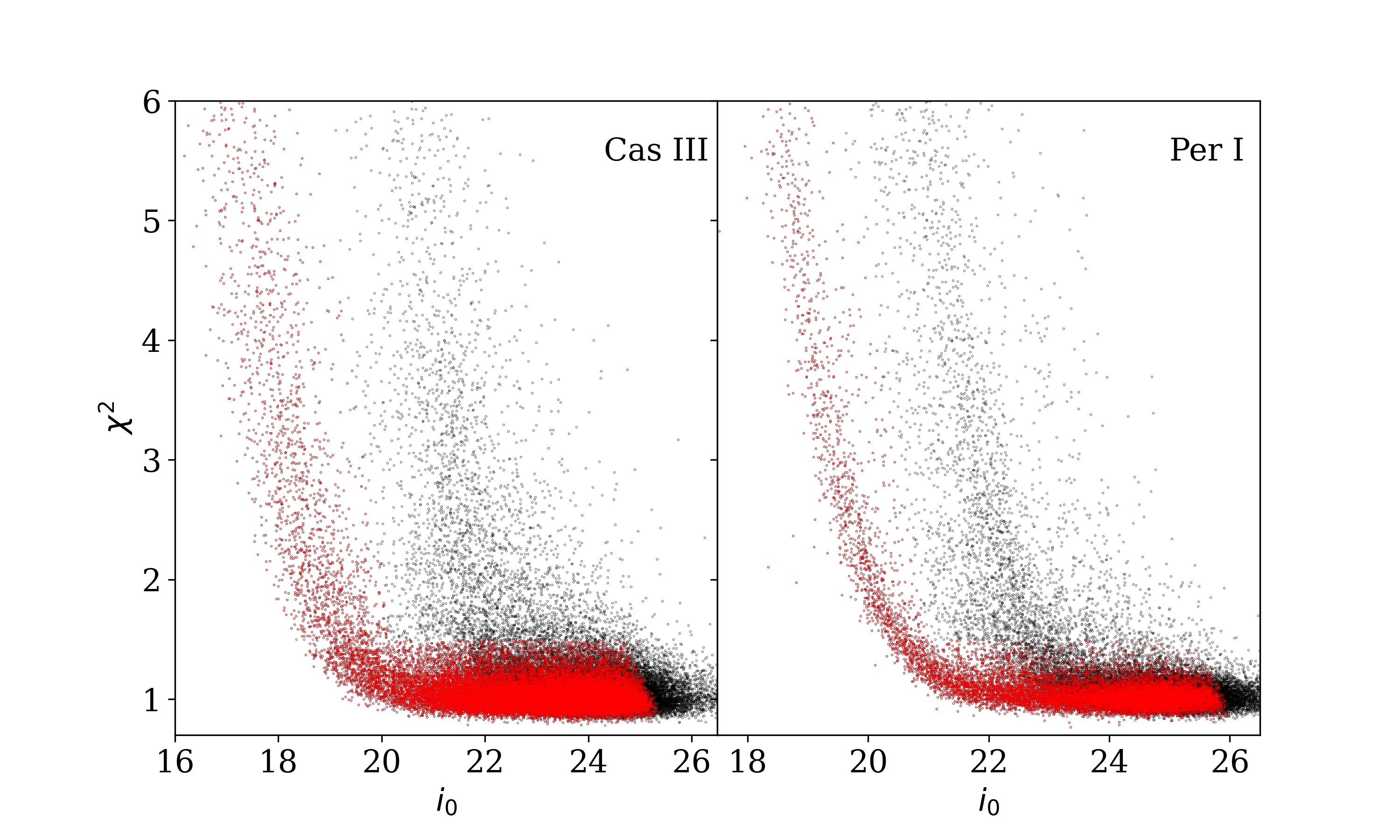}
\caption{Example of the $\chi^2$ selection criteria applied to the source lists for Cas~III and Per~I.  The $\chi^2$ value from DAOPHOT is plotted vs. $i_0$ magnitude; objects that we select are shown in red and those in black are discarded.}
\label{fig:chi}
\end{figure*}

\begin{figure*}[h!]
\plotone{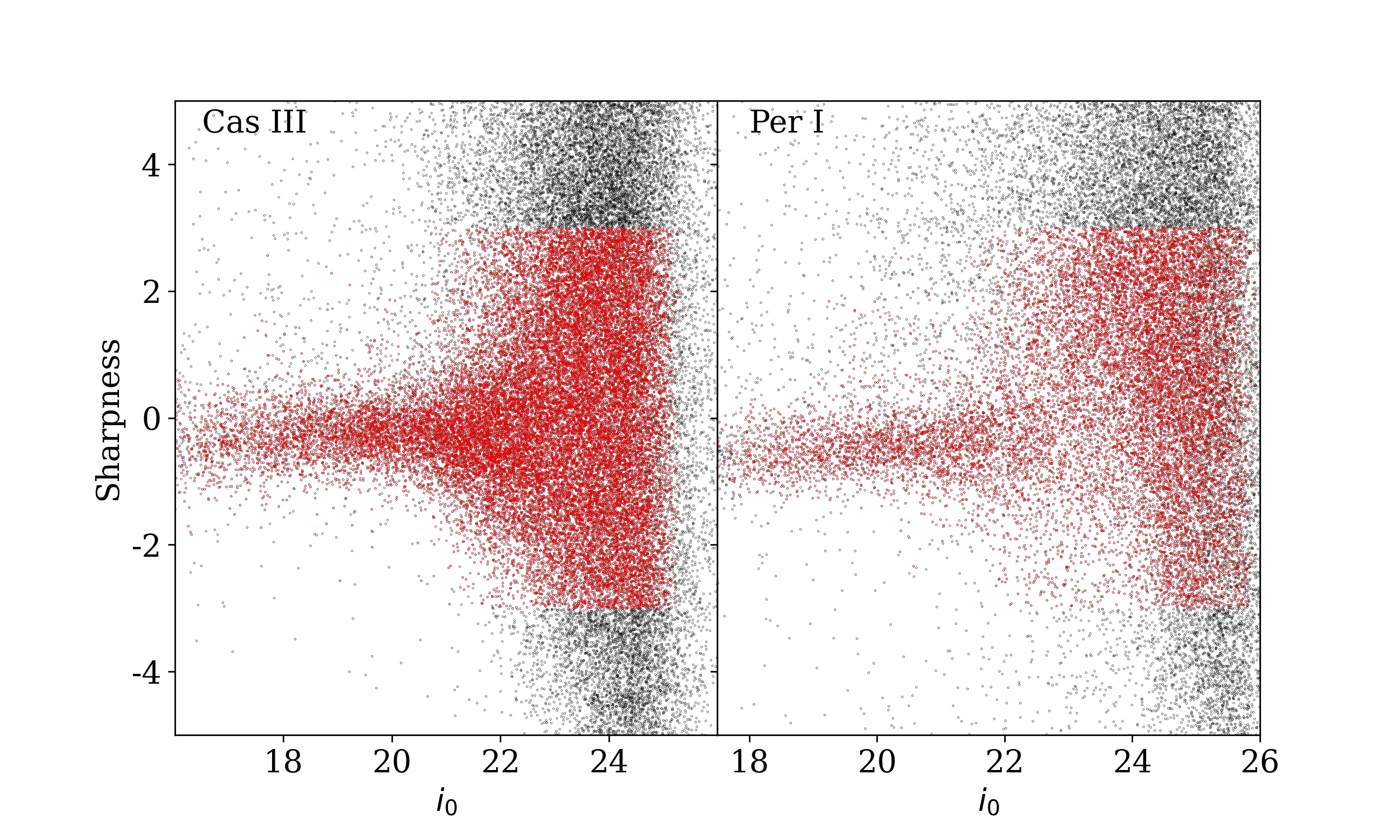}
\caption{Example of the selection criteria applied to the DAOPHOT photometry lists for Cas~III and Per~I.  The sharp parameter is shown as a function of calibrated $i_0$ magnitude; selected objects are plotted in red and rejected ones appear in black.}
\label{fig:sharp}
\end{figure*}

To detect and measure the sources in the final combined $g$- and $i$-band images of Cas~III and Per~I, we carried out source detection and PSF-fitting photometry with  DAOPHOT and ALLFRAME \citep{stetson87,stetson94}. We first constructed a model PSF using a set of $\sim$300 stars in each image that were bright and unsaturated. All sources within the images that had peak counts greater than or equal to three times the standard deviation of the sky background level were then fitted with the model PSF, and the magnitudes of each of these sources were calculated.  The source lists were first culled to remove faint stars with large photometric uncertainties ($>$0.4~mag).
We then further limited the source lists based on the value of the goodness-of-fit parameter $\chi^2$ from the PSF fitting. At faint magnitudes ($i_0$ between $\sim$20 and 25 mag), we retained sources with 
$\chi^2$ values less than 1.5, whereas we allowed a wider range of values for brighter sources. Figure \ref{fig:chi} illustrates 
the source selection in the $i$-band filter for both galaxies. We also restricted the source list to objects with
$|$sharp$|$ $<$ 3 (see Figure~\ref{fig:sharp}), to remove contaminants such as cosmic rays and background galaxies.
The final source catalogs for the Cas~III and Per~I images included 28,923 stars and 15,263 stars, respectively.  The photometric catalogs for the stars in the two galaxies are presented in Tables~\ref{tab:phot.casiii} and \ref{tab:phot.peri}. 
The catalog includes a sequence number for each star, the star's Right Ascension and Declination, the calibrated, dereddened $g_0$ and $i_0$ magnitudes and associated instrumental errors, and the calculated Galactic absorption values ($A_g$ and $A_i$) that were applied.

We carried out a series of artificial star experiments in order to quantify the detection completeness of the data and the uncertainties in our photometric measurements. We injected a total of $\sim$700,000 artificial stars into the Per~I images and $\sim$800,000 stars into the Cas~III images, which were slightly larger in size compared to the Per~I images. 
We carried out 20 different experiments for each set of images, in order to limit the effect of crowding on the overall detection process. In each experiment, we distributed the stars evenly across the frame and gave them colors and magnitudes within the ranges defined by the actual stars in our photometric samples. We carried out the same processing steps on the images that included the artificial stars as were carried out on the original images, and then measured the magnitudes of the injected stars.  The results of these experiments indicate that the 50\% color-averaged completeness levels are $g$ $=$ 24.4 and $i$ $=$ 23.6 in the Cas~III images and $g$ $=$ 25.2 and $i$ $=$ 24.3 in the Per~I images.  The 50\% completeness occurs at slightly brighter magnitudes inside the region within one half-light radius ($\sim$4.7\arcmin\ for Cas~III and $\sim$1.4$\arcmin$ for Per~I; see Sec.~\ref{sec: structural}) because of stellar crowding in the inner parts of the galaxies; the values become $g$ $=$ 24.3 and $i$ $=$ 23.3 in the Cas~III images and $g$ $=$ 25.1 and $i$ $=$ 24.1 in the Per~I images. Based on the artificial star experiments, our photometric errors are approximately 0.1 mag at $g$ $=$ 24.6 and $i$ $=$ 23.6 in the Cas~III data and at $g$ $=$ 24.8 and $i$ $=$ 24.2 in the data for Per~I.

\section{Properties of the Galaxies}
\label{sec: properties}

\subsection{Color Magnitude Diagrams and Stellar Spatial Distributions}
\label{sec: CMD}

\begin{figure*}[h!]
\plotone{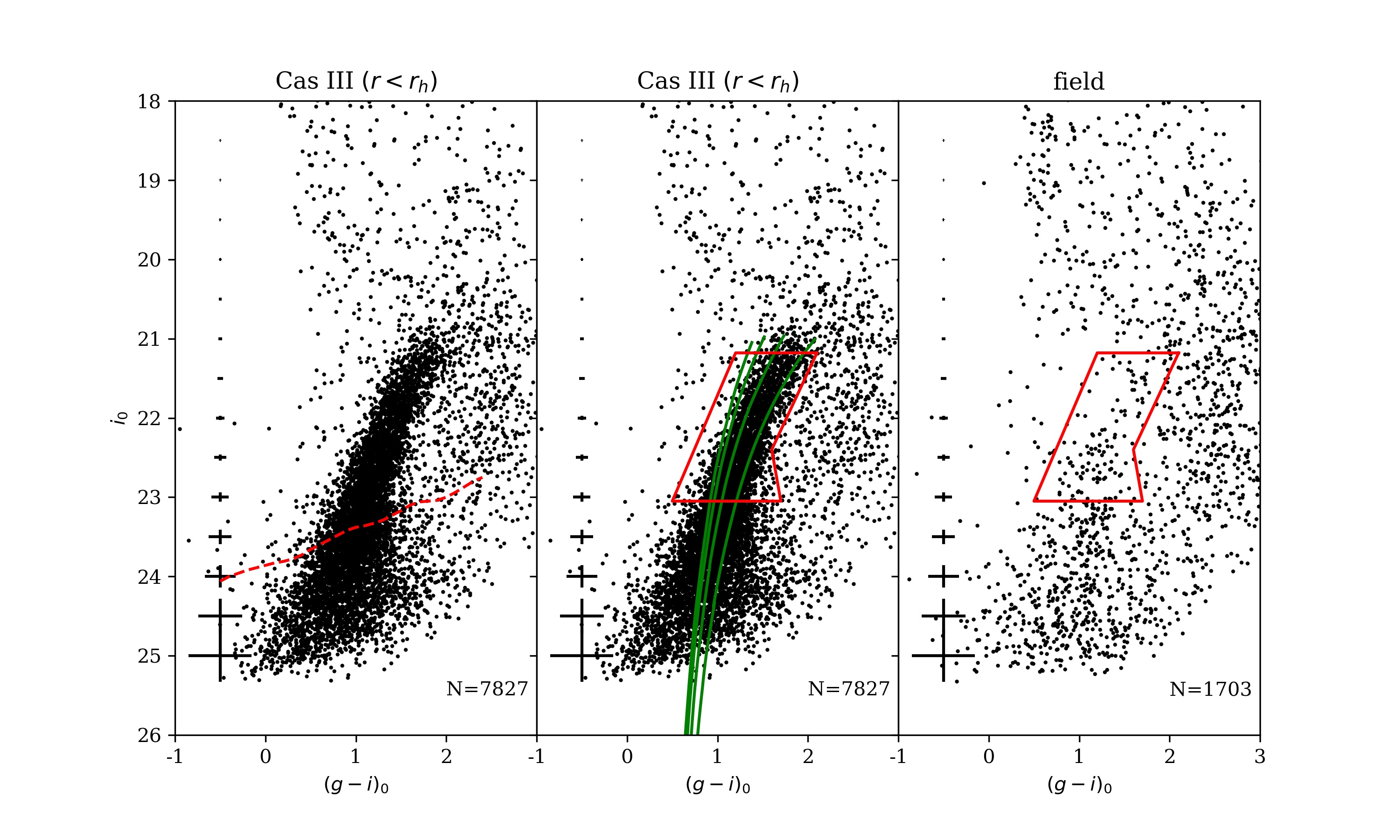}
\caption{The left and middle panels show the dereddened CMD of all stars within one half-light radius from the center of Cas~III. The 50\% completeness curve is marked with a dashed line in the left panel and the typical photometric uncertainties at each magnitude are plotted in all three panels.  Isochrones from \citet{dotter08} for an age of 12~Gyr and [Fe/H] values of -2.5, -2.0, -1.5, and -1.0 are marked (green lines) in the middle panel. The middle and right panels show the RGB selection box (red polygon). The right panel shows the CMD of stars within the rectangular background regions marked in Figure~\ref{fig:spatial.casiii}, with the numbers of sources rescaled so that the CMDs sample the same areas (see Sec.~\ref{sec: CMD} for details).}
\label{fig:cmd.casiii}
\end{figure*}

\begin{figure*}[h!]
\plotone{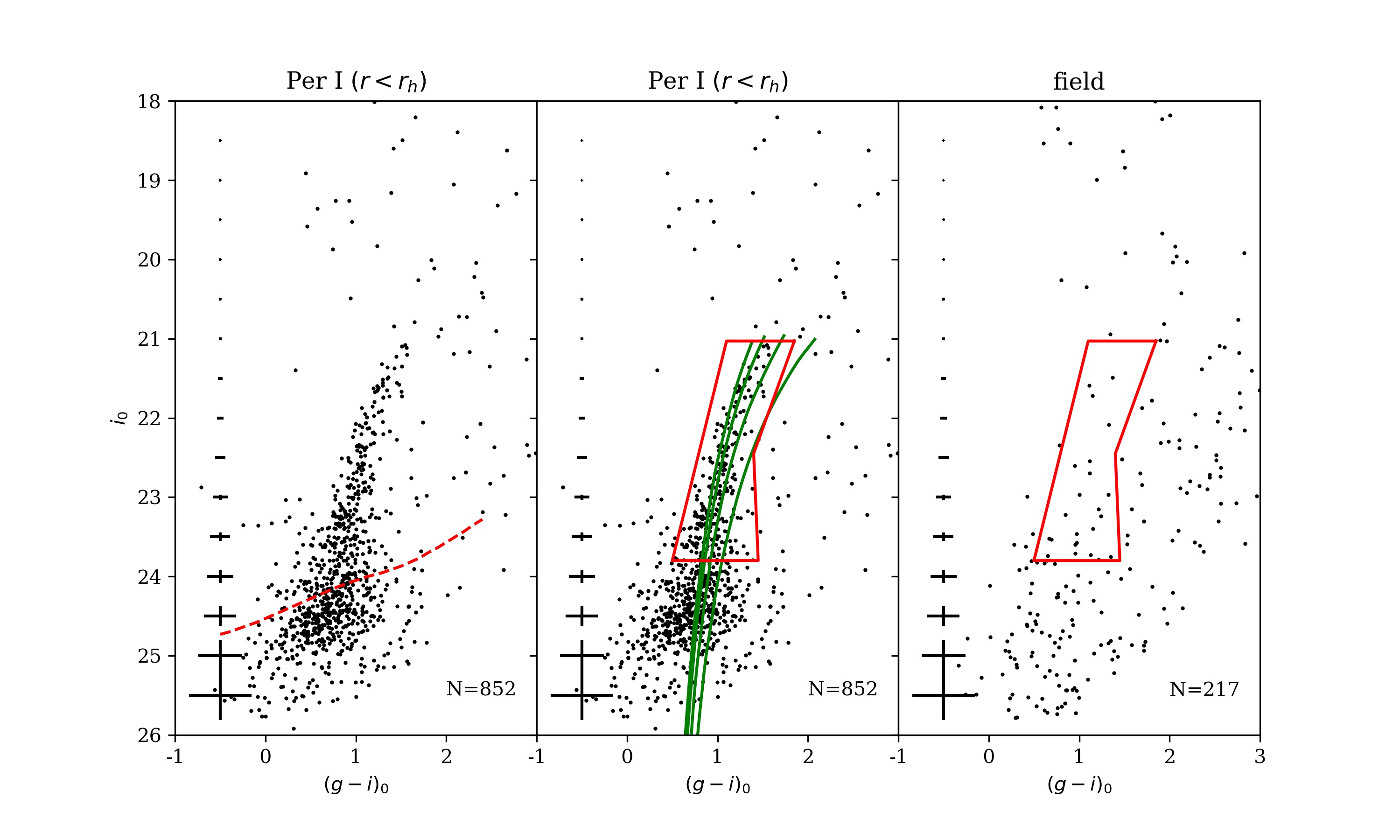}
\caption{The left and middle panels show the dereddened CMD of all stars within one half-light radius of the center of Per~I.  The right panel shows the CMD of stars in the rectangular background regions marked in Figure~\ref{fig:spatial.peri}. The 50\% completeness level, photometric uncertainties, RGB selection box, and isochrones are plotted on the figures in the same way as in Figure~\ref{fig:cmd.casiii}.}
\label{fig:cmd.peri}
\end{figure*}

\begin{figure*}[h!]
\plotone{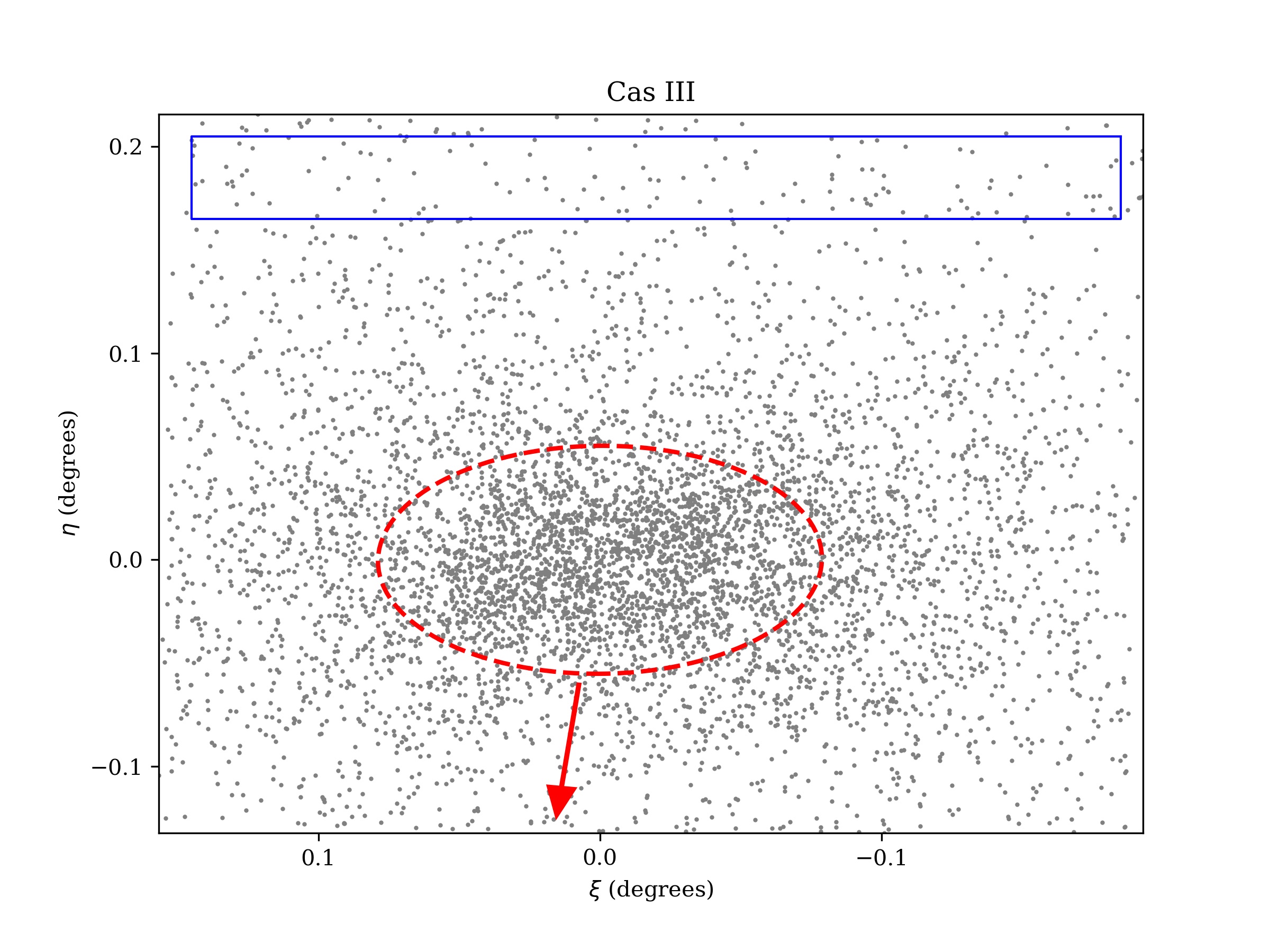}
\caption{The spatial locations across the WIYN pODI field-of-view of the 5725 RGB star candidates selected from Cas~III's CMD (Fig.~\ref{fig:cmd.casiii}). The positions of the stars are plotted in standard coordinates with respect to the galaxy center.  The half-light radius derived in Sec.~\ref{sec: structural} is marked with a dashed ellipse and the area of the image used to create the field CMD and measure the background density is shown as a blue rectangle. A red arrow indicates the direction to M31.}
\label{fig:spatial.casiii}
\end{figure*}

\begin{figure*}[h!]
\plotone{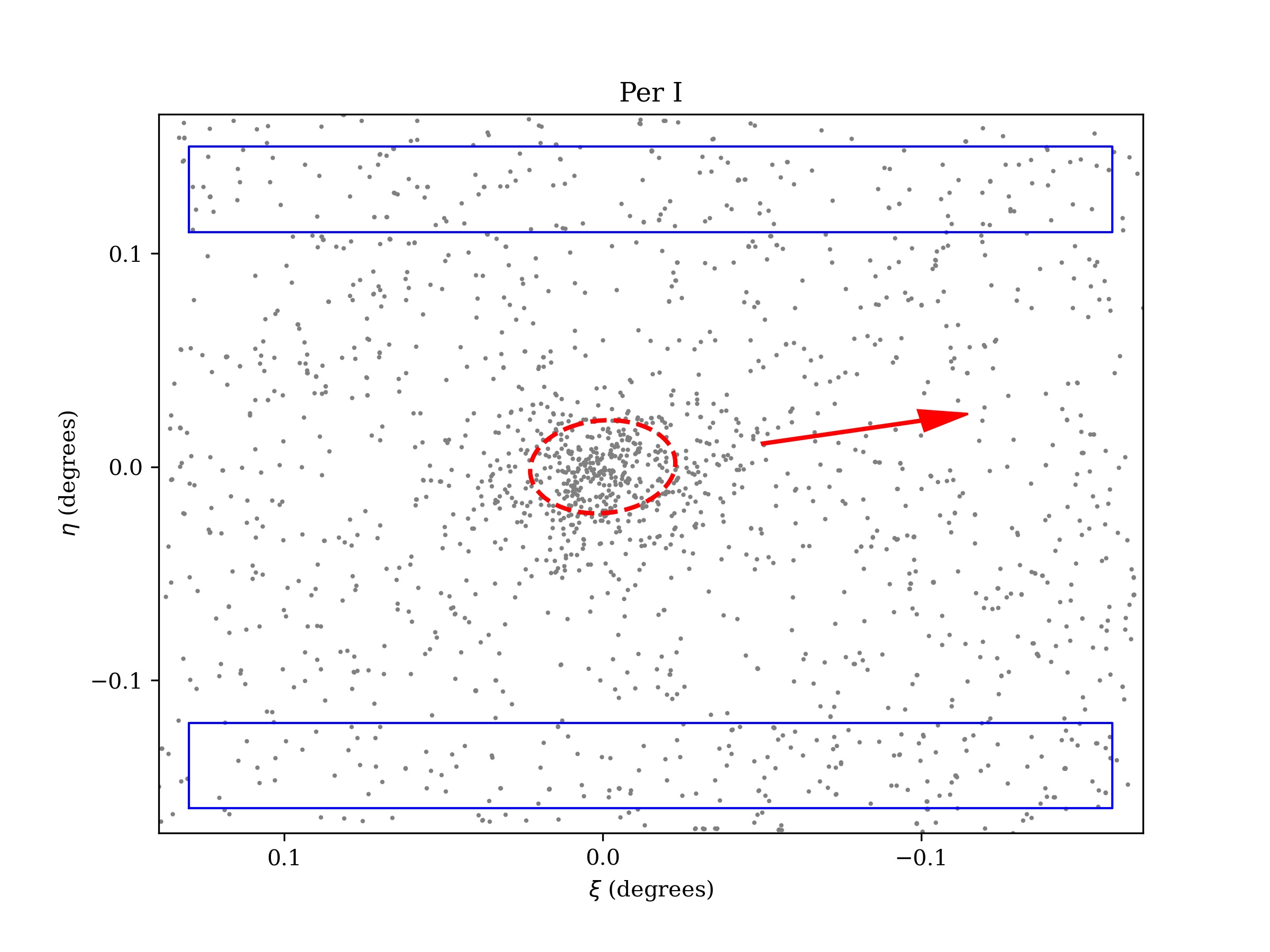}
\caption{The spatial locations across the WIYN pODI field-of-view of the 1817 RGB star candidates selected from Per~I's CMD (Fig.~\ref{fig:cmd.peri}).  The stars' positions are shown in standard coordinates with respect to the center of Per~I.  The dashed ellipse marks the half-light radius derived in Sec.~\ref{sec: structural} and the blue rectangles show the regions used to create the field CMD. The red arrow shows the direction to M31.}
\label{fig:spatial.peri}
\end{figure*}

The CMDs for the stars 
in Cas~III and Per~I are shown in Figures~\ref{fig:cmd.casiii} and \ref{fig:cmd.peri}, respectively. In both figures, the left and middle panels show the stars that are within one half-light radius of the galaxies' central coordinates (see Sec.~\ref{sec: structural} for a description of how we determined these 
parameters)
and the rightmost panels show the stars that appear in comparison regions near the edges of the images, beyond $\sim$2$-$3 half-light radii from the galaxy centers.
The locations of the comparison field regions are marked in Figures \ref{fig:spatial.casiii} and \ref{fig:spatial.peri}. 
The areas of the images used to produce the galaxy CMDs 
were smaller than the areas of the comparison field regions, so  
to construct the field CMDs 
we calculated the ratio of the areas of the galaxy and field regions and then randomly selected the appropriate proportion of stars from the field regions to ensure that we were sampling equivalent areas of sky.  The photometry shown in the CMDs has been corrected for Galactic reddening, as described in Sec.~\ref{sec: photometry}. The photometric uncertainties and the 50\% completeness limits estimated from the artificial star experiments are included in the figures.

The CMDs for both the Cas~III and Per~I fields show a clear sequence of Red Giant Branch (RGB) stars that is traced down to the faintest magnitudes in our images.  Our photometry with WIYN reaches approximately 2$-$3 magnitudes fainter than the Pan-STARRS photometric data in the discovery papers for the two galaxies \citep{martin13a,martin13b}.  
Given the distances of the two galaxies (see Sec.~\ref{sec: trgb}), the HB stars in Per~I and Cas~III would be expected to appear at magnitudes of $i_0$ $>$24,
and indeed this is what the HST studies of the galaxies that reached much greater depth than our study found \citep{martin17a,weisz19a}. This is below our 50\% completeness level and therefore the HB is not distinguishable in 
the CMD of either galaxy.
Neither CMD
shows any other obvious features.
We have overlaid isochrones from \citet{dotter08} that were calculated for an old stellar population (12~Gyr)
with a range of metallicity values ([Fe/H] $=$ $-$1.0 to to $-$2.5). 
As mentioned in the Introduction, \citet{martin14} estimated [Fe/H] $=$ $-1.7\pm0.1$ for Cas~III and [Fe/H] $=$ $-2.0\pm0.2$ for Per~I from spectroscopy of a sample of RGB stars in each galaxy, and the RGB sequences in our CMDs relative to the overlaid isochrones appear to be generally consistent with those measurements.
The RGB of Cas~III appears relatively broad compared to the same feature in Per~I and in the CMD for Lac~I \citep{rhode17}.  This seems to suggest that Cas~III  could have multiple stellar populations present, rather than being dominated by stars with a single age and metal abundance.  It is worth noting that \citet{martin14} identified a "handful" of carbon stars in the spectra they obtained for stars in both Cas~III and Lac~I, and concluded that the presence of these stars could suggest that these two galaxies host intermediate age populations.

The portions of the CMDs that were used to select the galaxies' RGB star populations are marked with red solid lines in Figures~\ref{fig:cmd.casiii} and \ref{fig:cmd.peri}.  We constructed the selection boxes so that they would cover the full color spread of the RGB and include stars with brightnesses ranging from the TRGB down to the magnitudes where the detection completeness began to decrease and the photometric uncertainties became large ($i_0$ $\sim$23 for Cas~III and $i_0$ $\sim$23.8 for Per~I). A total of 5725 stars lie within the RGB selection box in the Cas~III field and the Per~I field has 1817 stars within the selection box. The field CMDs in the right-hand panels of the figures show that most of the contamination at the faint end of the CMD is due to unresolved background galaxies, and most of it occurs at magnitudes below the bottom edge of the selection boxes. 

The spatial distributions of the RGB candidates across the entire field 
are displayed in Figures~\ref{fig:spatial.casiii} and \ref{fig:spatial.peri}. 
Our derived half-light radii (see Sec.~\ref{sec: structural}) are marked in the spatial distribution figures with a dashed line and the direction to M31 is designated with an arrow.  Both Cas~III and Per~I appear generally elliptical in shape and are clearly visible as overdensities of stars in the full-field spatial distribution plots.  Although the two galaxies have similar distances (see Sec.~\ref{sec: trgb}), Cas~III has a much larger extent on the sky than Per~I, filling much of the pODI frame and extending to the edge of the southern side of the image (the galaxy is slightly offset toward that side of the image).  In order to construct the comparison field for Cas~III to use for the field CMD shown in the rightmost panel of Figure~\ref{fig:cmd.casiii}, we positioned a rectangle on the north side of the frame that was located just beyond the 3~r$_h$ limit.  For Per~I, we were able to choose two rectangular regions near the edges of the image on either side of the galaxy, well beyond 3~r$_h$, as our comparison fields.  We used these same rectangular regions to provide an initial estimate of the background surface density (the surface density of contaminating objects) for the structural parameter fitting process described in Section~\ref{sec: structural}.

\subsection{Distance Estimates from the Tip of the Red Giant Branch Method}
\label{sec: trgb}

The TRGB distance method 
makes use of the fact that low-mass RGB 
stars 
that are undergoing the Helium flash have a specific luminosity that 
varies by only $\sim$0.1 mag; this is true for stars with a wide range of ages but the same metallicity  \citep{dacosta90,lee93}.  
Therefore the location of the TRGB makes a useful standard candle in star clusters and galaxies with resolved stellar populations, if one can successfully disentangle the low-mass stars that are genuinely undergoing the Helium flash from stars with different masses and/or evolutionary stages \citep{lee93}.

We used our pODI photometry to derive TRGB distances for Cas~III and Per~I by following 
the same method as in our earlier study of Lac~I \citep{rhode17}, which is 
based on the techniques presented in \citet{makarov06} and modified by \citet{wu14}. We began by constructing a model luminosity function for each data set and convolved it with the information about completeness and uncertainty that 
came from our artificial star tests.  
We then constructed an observed luminosity function for each galaxy, using stars within two half-light radii of the galaxy center. (See \citet{rhode17} for an example of a typical luminosity function constructed in this way.)
We constrained the stellar samples to stars with $g-i$ colors in the range 1.2$-$1.8, in order to select the metal-poor population and to limit the contribution from foreground Milky Way stars. We then fitted the model luminosity functions to the observed ones, which allowed us to determine the location of the discontinuity in the observed functions that marks the location of the TRGB.

Combining the observed apparent magnitude of the discontinuity in the luminosity function with the expected absolute magnitude 
yields the distance modulus for the galaxy. The absolute M$_i$ magnitude of the TRGB is essentially constant at $-$3.44$\pm$0.1 mag in stellar populations with [Fe/H] in the range $-$1.0 to $-$2.4 
\citep{bellazzini08}.
For Cas~III we find the apparent magnitude of the TRGB to be m$_i$ $=$ 21.18$\pm$0.07 mag, and the corresponding value for Per~I is m$_i$ $=$ 21.03$\pm$0.08 mag. These translate to distance modulus values of 24.62$\pm$0.12 mag for Cas~III and 24.47$\pm$0.13 mag for Per~I. The uncertainty on the magnitude of the observed TRGB is calculated by adjusting the magnitudes of the stars that go into the luminosity function according to their individual photometric uncertainties and then again finding the TRGB discontinuity; this process is repeated so that the full range of the measured stellar magnitudes is sampled. 
The distances to the galaxies given by their distance modulus values are 
$839^{+48}_{-45}$~kpc and $783^{+48}_{-45}$~kpc
for Cas~III and Per~I, respectively. 
Combining these distances with the distance to M31 
\citep[drawn from the catalog in][]{mcconnachie12}
and the angular distance on the sky between the galaxies yields a 3D distance from M31 of 156$^{+16}_{-13}$ kpc 
for Cas~III and 351$^{+17}_{-16}$ kpc for Per~I. 
Our measured TRGB distance to Per~I is entirely consistent with the distance estimated in the discovery paper \citep[$m-M$ $=$ 24.49$\pm$0.18;][]{martin13b}.  However, our distance to Cas~III is slightly larger than the discovery paper distance 
\citep[$m-M$ $=$ 24.45$\pm$0.14;][]{martin13a},
although it is consistent within the errors on the measurements.
Table~\ref{tab:properties all}
lists our estimated TRGB distances, along with the basic properties that we derive for the galaxies, including the total magnitude and color and some of the structural parameters discussed in Section~\ref{sec: structural}. In the same table, we have listed the properties of Lac~I, some of which are drawn from the \citet{rhode17} study and some from the analysis presented here, so that the three galaxies can be easily compared.

As mentioned in Section~\ref{sec: introduction}, more accurate distance estimates based on deep HST observations and photometry have recently been derived for many of the M31 satellites, including Lac~I, Cas~III, and Per~I \citep{weisz19a,savino22a}.  We list these various distances measurements in Table~\ref{tab:distances} in order to make it easier to compare them with our derived TRGB estimates. Our ground-based TRGB distance modulus measurements for all three galaxies are consistent with the corresponding HST-derived TRGB estimates from \citet{weisz19a}, given the estimated errors ($\sim$0.11-0.13~mag) on our values.  Our TRGB distance modulus values are systematically larger than the RR~Lyrae-derived values from \citet{savino22a} in all three cases, although the values for Lac~I and Cas~III still agree within our estimated errors.  \citet{savino22a} carry out a detailed exploration of the differences in distances derived from the TRGB and HB methods compared to those calculated from RR~Lyrae variables and find that TRGB distances can often be systematically overestimated because of sparsely populated RGBs, so our larger values are perhaps not unexpected.  In any case the general agreement between our TRGB distances and the corresponding HST TRGB values from \citet{weisz19a} serves as a useful check on the validity of our methods.
\citet{savino22a} make a strong case that the RR~Lyrae-derived distances are the most accurate of the various determinations for these galaxies, so for some of the distance-dependent quantities presented in this paper
(e.g., see Section~\ref{sec: structural luminosity}), 
we explicitly note the effect that using the RR~Lyrae distance would have on the final value.

\subsection{Structural Properties}
\label{sec: structural}

\subsubsection{Methods}
\label{sec: structural methods}

Following the general technique described in \citet{martin08}, we used Maximum Likelihood Estimation (MLE) to derive the structural parameters of
the stellar distributions of the dwarf galaxies. We began by assuming a Sersic model for the stellar distributions 
and adopted a likelihood function associated with that model \citep{graham05,martin08}.  We then used the "emcee" Python implementation of a Markov chain Monte Carlo (MCMC) algorithm \citep{foreman-mackey13a} to explore the parameter space and identify the best-fitting structural parameters that would maximize the likelihood function. The set of parameters we explored were the Sersic index ($n$), half-light radius (r$_h$), ellipticity, position angle, and central coordinates of the galaxies.  We fitted the spatial distribution of the stars using an exponential model ($n$ fixed at 1), as well as allowing the Sersic index to vary, in order to help facilitate our searches for substructure in the galaxies' stellar distributions (see Section~\ref{sec: substructure} for details).  

For the MCMC fitting process, we discarded the first 3,000 steps in the chain in order to begin the convergence process and minimize any bias associated with the starting positions of the parameters.  We allowed 30,000 more steps in the chain to ensure convergence
and took the median value of the distribution of each fitting parameter as the final best-fitting value.  The 16th and 84th percentile values of the distributions were used to define the uncertainties on the median values. 

To derive structural parameters for Cas~III and Per~I, we used the objects that fell within 
the red RGB selection boxes shown in the CMDs 
(Figures~\ref{fig:cmd.casiii} and \ref{fig:cmd.peri}). 
As mentioned, Figures~\ref{fig:spatial.casiii} and \ref{fig:spatial.peri} show the spatial distribution of the 5725 RGB candidates in the Cas~III field and the 1817 RGB candidates in the Per~I field. 
While the 
background surface density (i.e., the surface density of objects like foreground stars and unresolved background galaxies that remain in our RGB candidate samples) can be allowed to vary in the fitting process, we chose 
to estimate this value from the data 
in order to reduce the number of free parameters that needed to be determined via the MLE process.
We calculated the background surface density using the same rectangular regions that were used to construct the field CMDs  
shown in Figures~\ref{fig:cmd.casiii} and \ref{fig:cmd.peri}.
The background surface density of RGB candidates calculated in this way is
2.59$\pm$0.23 arcmin$^{-2}$ for Cas~III and 3.66$\pm$0.22 arcmin$^{-2}$ for Per~I. 

Because the stellar distribution of Cas~III covers much of the pODI field-of-view,
our selected background regions may include stars that actually belong to the galaxy; thus it is reasonable to assume that the background estimate for Cas~III may be higher than the true value.  In the case of Per~I, we found when we constructed the radial profile (see Fig.~\ref{fig:profile.peri}) that the background surface density was not uniform but instead showed significant variation at radii $\sim$5$-$10 arcmin from the galaxy center.  This suggested that our measurement of the stellar surface density within discrete rectangular regions at particular radii might not yield the best estimate of the background.  As a result, we decided to adjust the background surface density value based on the value that provided the best fit to the observed radial profile.  We adjusted the background surface density estimate between 3 and 3.75 arcmin$^{-2}$ in increments of 0.05 and used the MLE algorithm to find a new set of best-fitting parameters associated with that estimate. 
We then took the sum of the squared residuals (SSR) between the surface density values of our radial profile and the surface density estimated by our model at that radius. The
background surface density and the corresponding best-fitting parameters were chosen as the final set of parameters. 
The final background surface density we used for Per~I is 3.30~arcmin$^{-2}$.

The \citet{rhode17} study already includes structural parameters that were derived using the Lac~I photometry and an exponential model. 
Nevertheless, because we wished to search for evidence of substructure within the stellar distributions of all three M31 satellites imaged with WIYN, 
we carried out the complete fitting process for Lac~I along with the two other galaxies.  This provided a useful confirmation that our MLE technique was working correctly and allowed us to search for substructure using the same approach for all three galaxies. 
The RGB candidate stars in Lac~I were detected in the images and selected from the CMD using the same basic methods as described here for Cas~III and Per~I.  The RGB candidate star list used to derive the structural parameters for Lac~I includes 7010 stars with magnitudes down to $i$ $=$ 24, and we assumed the same background surface density as was used in \citet{rhode17}: 6.82~arcmin$^{-2}$.  We refer the reader to \citet{rhode17} for additional details about the Lac~I photometric data set and RGB star candidate selection.  

\subsubsection{Results}
\label{sec: structural results}

\begin{figure*}[h!]
\plotone{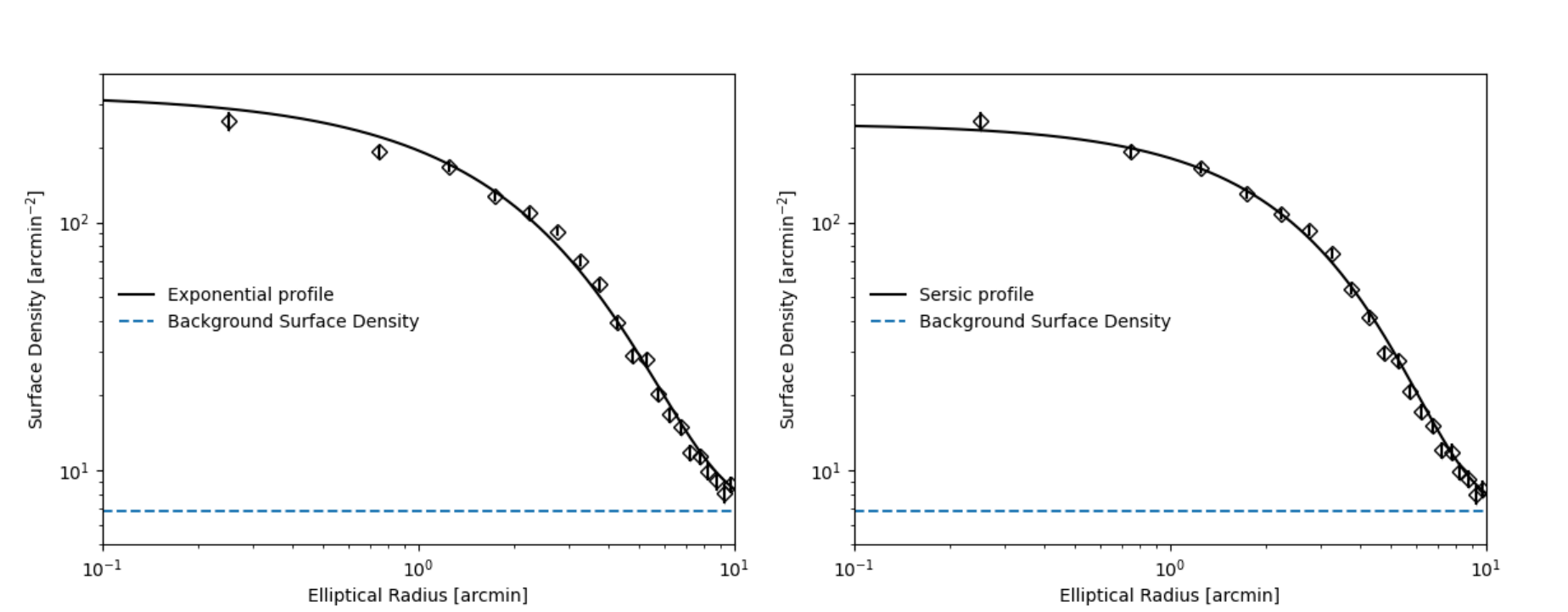}
\caption{The surface density of the RGB candidate stars in Lac~I as a function of elliptical radius. The solid black line is the best-fitting function derived from MLE process described in Section~\ref{sec: structural methods}; the left panel shows the exponential function and the right shows the Sersic function with $n=$0.83. Open diamonds indicate the surface density within each elliptical bin. The dashed line indicates the background surface density.}
\label{fig:profile.laci}
\end{figure*}

\begin{figure*}[h!]
\plotone{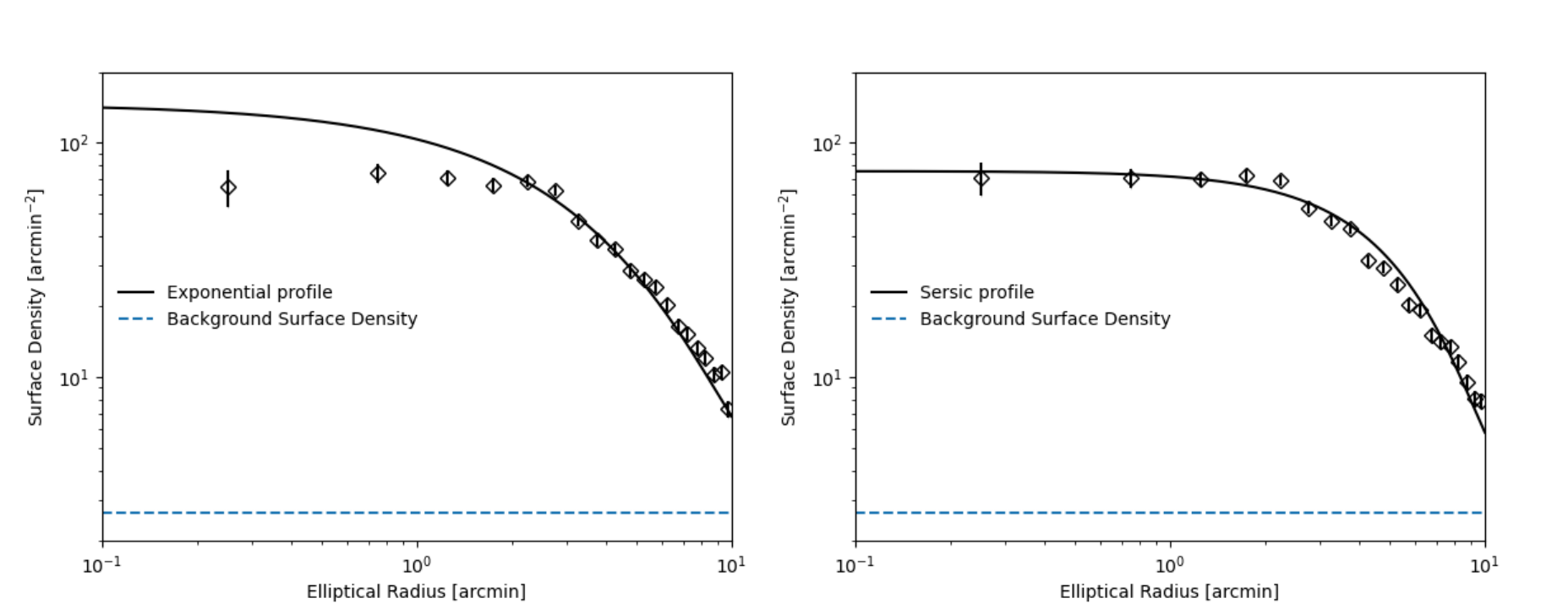}
\caption{Surface density of RGB candidates in Cas~III as a function of elliptical radius, plotted in the same way as in Figure~\ref{fig:profile.laci}. The exponential fit with $n=$1 is shown on the left and on the right is the Sersic function fit, with $n=$0.57.
}
\label{fig:profile.casiii}
\end{figure*}

\begin{figure*}[h!]
\plotone{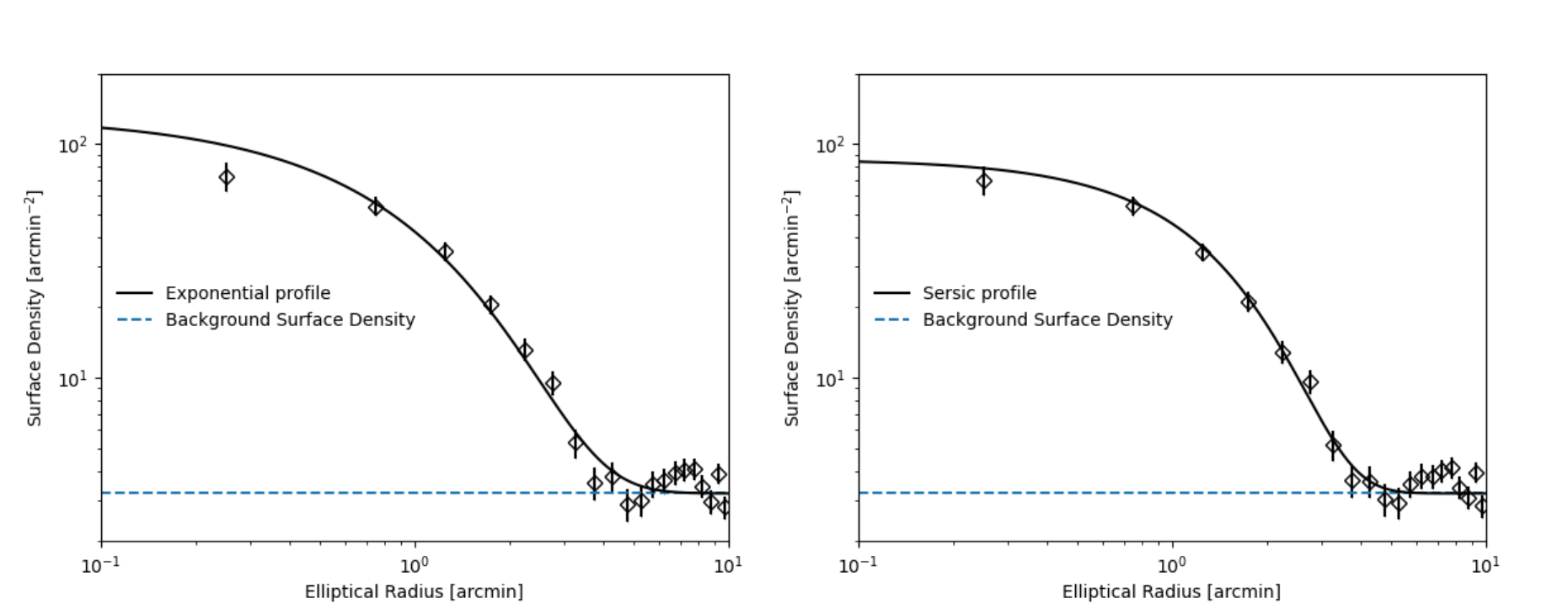}
\caption{Surface density vs. elliptical radius for the RGB candidate stars in Per~I, plotted in the same way as in Figure~\ref{fig:profile.laci}. The best-fitting exponential function appears in the left panel and the best-fitting Sersic function with 
 $n=$0.7
appears in the right panel.}
\label{fig:profile.peri}
\end{figure*}

The full results from the structural parameter fitting process are shown in Table~\ref{table:structural params all}, which lists the best-fitting parameters for both the exponential profile and the general Sersic profile. 
Figures~\ref{fig:profile.laci}$-$\ref{fig:profile.peri} show the one-dimensional versions of the stellar distributions -- i.e., the surface density of stars vs. elliptical radius 
-- for Lac~I, Cas~III, and Per~I.  The elliptical radius is calculated using the definition given in \citet{martin08} (their Equation 4). The left panel in each figure shows the exponential fit and the right panel shows the Sersic function fit.
Although the figures show the binned data points, we note that the MLE fitting process is carried out with the unbinned distributions.  We also note that because the MLE process yields slightly different central coordinates for the Sersic model and the exponential model (see Table~\ref{table:structural params all}), the exact values of the binned data points differ slightly for the two cases; for this reason, we show the data and best-fitting functions side-by-side rather than plotting them on top of each other.

In the case of the Lac~I surface density profile (Figure~\ref{fig:profile.laci}), both the exponential and Sersic functions closely mirror the data beyond $\sim$1$\arcmin$, departing only slightly from the observed data points in the innermost region of the galaxy.  The exponential fit overestimates the central surface density by $\sim$1-$\sigma$ at those radii, whereas the Sersic fit (with $n=$0.83) underestimates the fit there. This may be caused by stellar crowding and/or the presence of two bright Galactic foreground stars in that region; nevertheless both functions are generally representative of the observed profiles.

The model fits to the Cas~III data (Figure~\ref{fig:profile.casiii}) are notably worse by comparison. Most likely due to considerable stellar crowding in the central portion of the galaxy, neither model matches the data in the inner $\sim$3$\arcmin$.  The exponential model overestimates the data points inside $\sim$3$\arcmin$, matches the data fairly well from 3$-$8$\arcmin$, and then underestimates the data points further out.  The Sersic model with $n=$0.57 
is better able to reproduce the data points inside $\sim$3\arcmin\ and outside 8\arcmin, but overestimates the data points in between.  
The severe crowding in the central region of the galaxy leads us to mask out this area when we search for substructure in the stellar distribution (see Section~\ref{sec: substructure casiii}).

The fits to the Per~I radial profile data (Figure~\ref{fig:profile.peri}) are of reasonable quality for both functions. The most notable deviation occurs at 0.25$\arcmin$, where the observed radial profile appears to flatten relative to the exponential model.  The Sersic profile provides a better fit at all radii, suggesting that 
the exponential model may not be a good descriptor of the RGB star distribution in Per~I.

We can compare some of our derived structural parameters (Table~\ref{table:structural params all}) with results from previous work.  
The simplest comparison is for Lac~I:  as expected (since we are using the same set of RGB star candidates and a similar method), our best-fitting parameters for the exponential model agree within the estimated errors with the corresponding parameters in \citet{rhode17} in all cases.  
These parameters were, in turn, in general agreement with the corresponding parameters in the \citet{martin13a} discovery paper, although the \citet{rhode17} half-light radius was smaller and slightly more than 1$-\sigma$ discrepant with the \citet{martin13a} value.
The most recently published structural parameters for Per~I were derived by \citet{higgs21} using a Sersic function and an MLE method.  Comparing the results from our best-fitting Sersic function to their results, we find that our derived half-light radius and Sersic index ($\sim$1.4$\arcmin$ and $n=$0.7, respectively) agree with their values, but with substantially smaller uncertainties. 
We find a similar, quite small, ellipticity:  0.09$\pm$0.06 for our value compared to 0.04$\pm$0.08 for \citet{higgs21}.  Our derived position angle agrees with the value in \citet{higgs21} when the error bars of both values are taken into account; the nearly circular shape 
of the galaxy's stellar distribution is likely the reason that the position angle is not better determined in either study. 

For Cas~III, we can compare our exponential model fitting results 
with those from the \citet{martin13a} discovery paper.  In this case we find significant differences between our results and theirs. Fitting an $n=$1 distribution to our Cas~III data set yields a half-light radius of 4.7$\pm$0.09 arcmin, which is smaller than the 6.5~${+1.2 \atop -1.0}$ arcmin value from \citet{martin13a}.  Our derived ellipticity of  0.35$\pm$0.01 also differs from their value of 0.50$\pm$0.09. 

The differences between our structural parameters for Cas~III and those of the \citet{martin13a} paper may arise from the fact that Cas~III's stellar distribution fills a large proportion of the field-of-view of our images. \citet{munoz12} used simulations to investigate the observational requirements for the robust determination of structural parameters of the 
stellar distributions of dwarf galaxies.  They concluded that three criteria must be satisfied in order for 
parameters within 10\% of their true values to be derived: (1) the number of sample stars must exceed 1000; (2) the ratio of stellar surface density in the central vs. the background regions must be $>$20; and (3) the field of view must be at least three times the size of the galaxies' half-light radius.   Our Per~I data set satisfies all three of the \citet{munoz12} criteria, but our Cas~III data set satisfies only the first two because the radial coverage of our images does not quite reach 3 times the expected half-light radius of $\sim$6.5$\arcmin$. 

To test our assumption that inadequate spatial coverage was impacting our derived structural parameters for Cas~III, we generated a set of artificial data with an extent and spatial distribution matching those reported by \citet{martin13a} distributed across a field-of-view that meets the \citet{munoz12} criteria. Running the MLE algorithm on this data set yielded structural parameters that agree with those in the \citet{martin13a} discovery paper.  We then generated a second artificial data set,
again matching the \citet{martin13a} spatial distribution but with the same number of objects and falling within the same field-of-view of our Cas~III data.  Using this second data set as input to the MLE algorithm yielded a half-light radius of 4.6$\pm$0.07$\arcmin$ and an ellipticity of 0.40$\pm$0.01.  
The results of these tests confirm that the field-of-view and other characteristics of the data set have a significant impact on the derived structural parameters, and seem to explain at least in part the differences between our structural parameters for Cas~III and those in the discovery paper. 

\subsection{Total Magnitude, Central Surface Brightness, and Neutral Gas Content}
\label{sec: structural luminosity}

We calculated total absolute magnitude values for Cas~III and Per~I by following the same method used for our study of Lac~I and described in \citet{rhode17} and \citet{sand12a}. We started by adding the flux enclosed within one half-light radius in each galaxy, based on the sample of stars that lie within the RGB selection box shown in the center panel of Figures~\ref{fig:cmd.casiii} and \ref{fig:cmd.peri}.  To account for the flux contributed by foreground and background sources, we corrected the flux values using the stars that fall within the rectangular background regions marked in Figures~\ref{fig:spatial.casiii} and \ref{fig:spatial.peri}, after appropriately rescaling the regions 
so that RGB stars and foreground/background objects are sampled from regions of equal area.
We then doubled the corrected flux to account for stars in the portion of the galaxy outside one half-light radius.  Finally, we used several PARSEC luminosity functions \citep{bressan12a} with ages of $\sim$10$-$13~Gyr and metallicities [Fe/H] in the range $-$1.5 to $-$2.0 to account for the stars below our photometric detection limits.  The total absolute magnitudes and colors 
that we derived in this way are $M_g$ $=$ $-$12.1$\pm$0.1 mag and $(g-i)$ $=$ 1.4$\pm$0.3 mag for Cas~III and $M_g$ $=$ $-$8.8$\pm$0.3 mag and $(g-i)$ $=$ 1.0$\pm$0.3 mag for Per~I. Applying the filter transformation from \citet{veljanoski13a} yields an absolute $V$-band magnitude $M_V$ of $-$12.5$\pm$0.2 mag and 
$-$9.1$\pm$0.3 mag for Cas~III and Per~I, respectively. 
To derive the central surface brightness in the $V$-band, we use these total $M_V$ values plus the best-fitting exponential profiles for each galaxy; the value for Cas~III is $\mu_{V,0}$ $=$ 24.9$\pm$0.3~mag~arcsec$^{-2}$ and 
for Per~I is $\mu_{V,0}$ $=$ 25.7$\pm$0.3~mag~arcsec$^{-2}$.

In the Cas~III discovery paper, \citet{martin13a} estimated a total magnitude of $M_V$ $=$ 
$-$12.3$\pm$0.7 mag, which is consistent with our measured value but has a substantially larger error.  The discovery paper for Per~I \citep{martin13b}, which was based on relatively shallow photometry compared to our study, gives $M_V$ $=$ $-$10.3$\pm$0.7 mag, which is just under 2-$\sigma$ different than our estimated total $M_V$ magnitude.  We can also compare our total magnitude to that given in the \citet{higgs21} study:  they estimate $M_i$ $=$ $-$9.28 (with no error explicitly given)
and $g-i$ $=$ 1.13$\pm$0.43 mag, while we find $M_i$ $=$ $-$9.8$\pm$0.4 mag, and a $g-i$ color that is consistent with theirs but with a slightly smaller error. It is worth noting that, because the half-light radius we derived is smaller than that derived in the \citet{martin13a} paper (see discussion in Section~\ref{sec: structural results}), the total magnitude we calculate will also be fainter, so this likely accounts for some of the differences compared to total magnitude measurements from other studies.

We should also note that all of the absolute magnitudes we derive from our WIYN photometric data are calculated using the TRGB distances described in Section~\ref{sec: trgb}.  When we instead take the distances derived from the RR~Lyrae variable star study in \citet{savino22a}, our absolute magnitudes for Cas~III and Per~I change to $M_V$ $=$ $-$12.40 mag and $-$8.87 mag, respectively.  Our $i$-band absolute magnitude for Per~I becomes $-$9.6$\pm$0.4 mag, which makes it consistent with the value in \citet{higgs21} within our estimated uncertainty. 

\citet{martin14} used their spectroscopic observations to estimate
[Fe/H] values of $-$2.0$\pm$0.1, $-$1.7$\pm$0.1, and $-$2.0$\pm$0.2 for Lac~I, Cas~III, and Per~I,
respectively. They examined
the relationship between absolute $M_V$ magnitude taken from the \citet{martin13a,martin13b} discovery papers and the mean [Fe/H] from the stellar samples, and compared this to the luminosity-metallicity (L-Z) relation in \citet{kirby13}, derived from stellar spectroscopy of other Milky Way and M31 dwarf satellite galaxies.   They concluded that the properties of the three dwarf galaxies are consistent with those of other Local Group dwarf satellites, albeit with a slightly lower metallicity than expected; they noted that Lac~I deviated by a few dex from the \citet{kirby13} relation but that the uncertainties in [Fe/H] measurements, especially for M31 satellites, can be large. \citet{rhode17} estimated $M_V$ $=$ $-$11.4$\pm$0.3 mag, which brought Lac I slightly closer to its expected position in the L-Z plane. 

We can use our updated $M_V$ values for Cas~III and Per~I and the \citet{martin14} metallicities and compare them to the \citet{kirby13} L-Z relation (their Equation 3).  Given our estimated absolute magnitudes for Cas~III and Per~I of $-$12.5 mag and $-$9.1 mag, respectively, the \citet{kirby13} relation predicts an [Fe/H] value of $-$1.41 for Cas III and $-$1.80 for Per I.  The estimated metallicites for the galaxies from \citet{martin14} are within one or two times the RMS scatter in the \citet{kirby13} relation (0.16~dex), so the L-Z values of Cas~III and Per~I are fairly consistent with expectations.

\citet{martin14} also used their stellar velocity measurements to calculate the systemic line-of-sight velocities of Lac~I, Cas~III, and Per~I, and the velocity dispersions $\sigma_{vr}$ of the first two galaxies ($\sigma_{vr}$ $=$ 10.3$\pm$0.9~\kms\ for Lac~I and $\sigma_{vr}$ $=$ 8.4$\pm$0.6~\kms\ for Cas~III). For Per~I, they had velocities for fewer stars and could only come up with a "favored" velocity dispersion value of 4.2~\kms.  They then combined the measured velocity dispersions, sizes, and absolute magnitudes of Lac~I and Cas~III to estimate the half-mass radius of the two galaxies and compare this to the size-mass relation for Andromeda satellite galaxies derived by \citet{collins14}. They concluded that Lac~I and Cas~III have estimated sizes and masses that are consistent with expectations based on dark-matter-dominated models and observed values for other dwarf satellites around M31.  Although we find slightly different total magnitudes and half-light radii compared to previous work in some instances, our values are still generally within $\sim$1$-$2 times the quoted uncertatainty on the previous measurements, so our results do not change the general conclusions made by \citet{martin14} when they examined this.

Finally, we sought to compare the gas content of the galaxies to that of other dwarf satellite galaxies in the Local Group. Cas~III and Per~I are gas-poor dwarf spheroidal galaxies with HI-mass upper limits of M$_{\rm HI}$ = 7.6 x 10$^4$ \msun\ and 7.9 x 10$^4$ \msun, respectively, as derived by \citet{putman21} using data from the HI4PI survey \citep{hi4pi16}. We use these values to calculate upper limits on their gas richness using the newly-derived $V$-band luminosities in this work, 
finding values of M$_{\rm HI}$/L$_{\rm V}$ $<$ 
0.009 \msun/\lsun\ and 
0.22 \msun/\lsun\ for Cas~III and Per~I, respectively.  
Both ratios are in line with the corresponding values for the broader population of dwarf spheroidal galaxies around the Milky Way and M31 \citep{grcevich09,spekkens14,putman21}.

\begin{figure*}
\plottwo{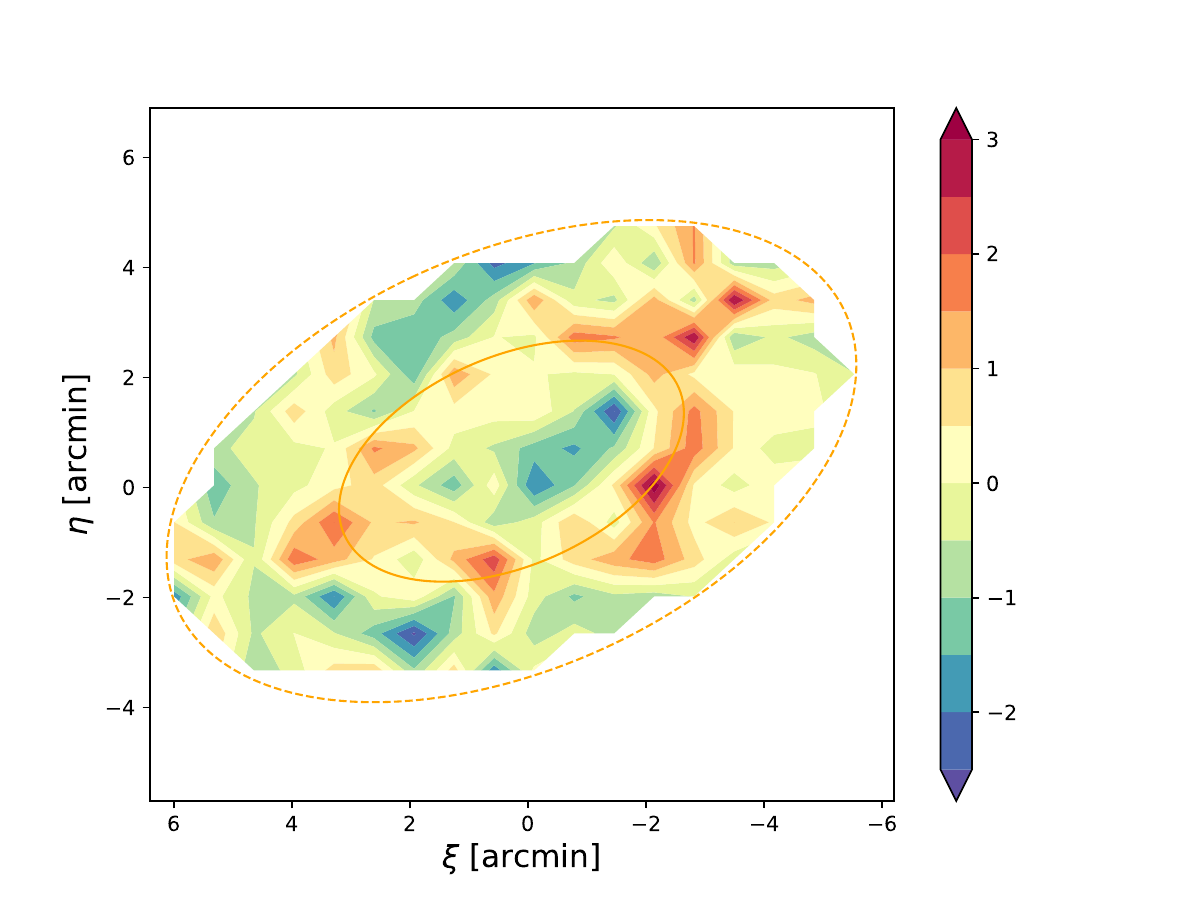}{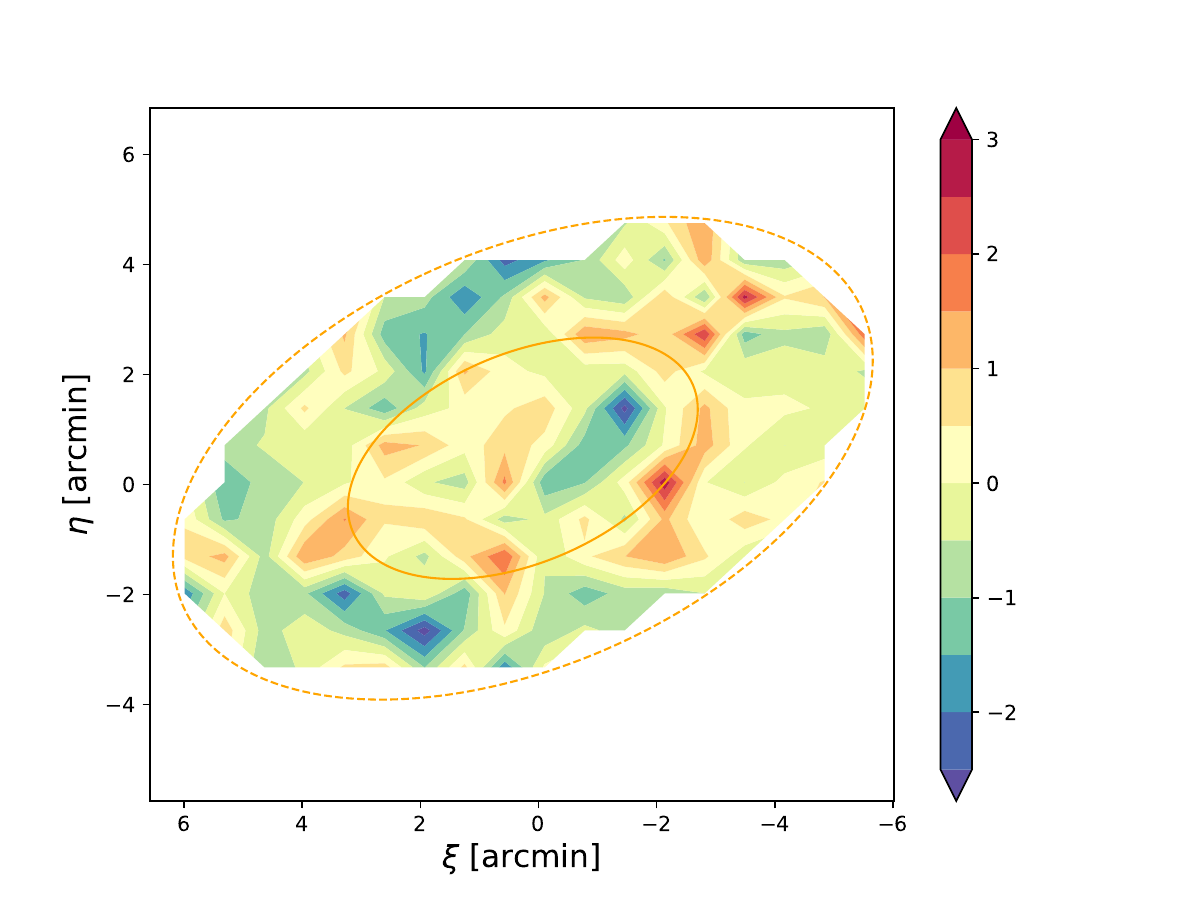}
\caption{
The residual maps for the inner two half-light radii of Lac I, with the comparison model as the best-fitting exponential model (left) and Sersic model 
(right).  The maps are generated by dividing the field into a 30x30-cell grid and calculating the estimated surface density of RGB stars at the center of each cell based on the best-fitting model.  The surface density is then converted to the number of objects expected within the cell, given the cell area.  The residual for a given cell is calculated by subtracting the expected number of stars within the cell from the actual number, and then dividing that quantity by the square root of the actual number of stars; see Section~\ref{sec: substructure methods} for more details.  Residual values are color-coded according to the color bar shown on the right, which ranges from approximately $+$3-$\sigma$ to $-$2.5-$\sigma$.  The solid and dashed ellipses mark one and two half-light radii, respectively. North is up and east is left.}
\label{fig:substructure.laci.inner}
\end{figure*}

\begin{figure*}
\plotone{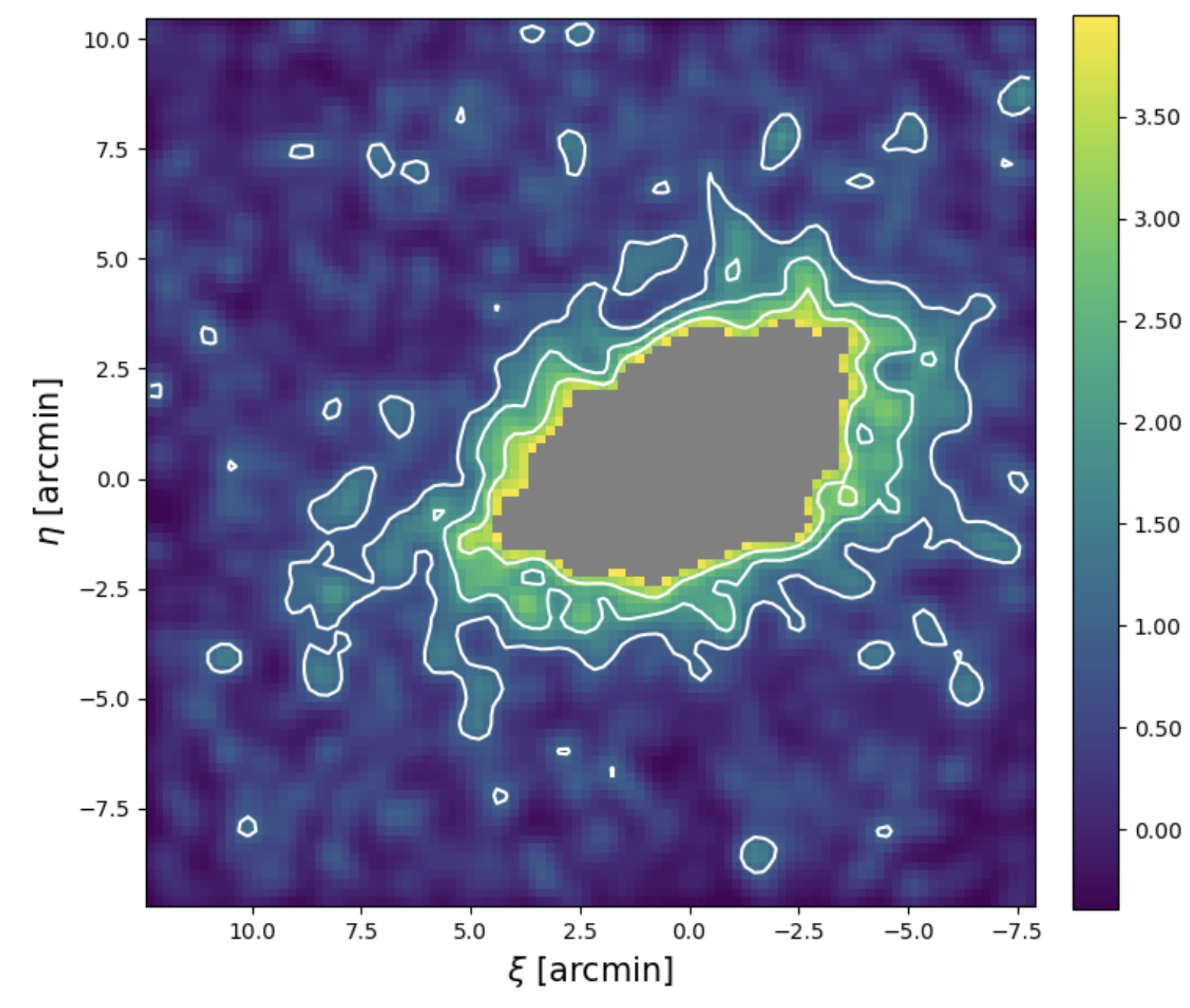}
\caption{
The map of the residuals in the region of the Lac~I field where the galaxy no longer dominates.  The map was generated as described in Section~\ref{sec: substructure methods}. Over-densities above 4-$\sigma$ are masked in gray and considered the main body of the galaxy. The white contours encompass 1-, 2-, and 3-$\sigma$ areas. 
North is up and east is to the left.}
\label{fig:substructure.laci.outer}
\end{figure*}

\section{Searching for Substructure in the Stellar Distributions of Lac~I, Cas~III, and Per~I}
\label{sec: substructure}

\subsection{Methods}
\label{sec: substructure methods}

After deriving best-fitting model parameters that describe the spatial distributions of the stars in each galaxy, we began searching for substructure within the RGB star populations.  The main objective was to examine whether over- or under-dense regions existed in the galaxies' stellar distributions that might provide clues about the galaxies' evolutionary histories.  

We began by dividing each field into a 30x30-cell grid and calculating the expected surface density at the center of the individual cells based on the best-fitting model. We then converted this surface density into the number of stars expected in each grid cell, using the area of the grid cell.  The expected number of stars in each cell provides an idealized comparison case (i.e., an unperturbed dwarf galaxy) and deviations from the model predictions are possible evidence that past events (e.g., accretion, merging, or other dynamical interactions) have perturbed the stellar system or deposited new stars into the galaxy. 

Next, we binned the stellar data into the same 30x30-cell grid and subtracted the expected number of objects in each grid cell from the actual number of RGB stars in the cell.  
The uncertainty in our result is governed by Poisson statistics so we determined the significance of a given residual by dividing the difference by the square root of the number of RGB stars in the cell.  
We carried out this process for both the best-fitting exponential model and general Sersic model.

We should note that the method described above works well for probing
the inner regions of the galaxies, where stellar densities are high and individual grid cells can be expected to contain tens of objects.
In the sparsely-populated outer regions of the galaxy, 
the relative uncertainties become so large that it is difficult to draw definite conclusions.  For this reason, we restrict ourselves to the inner two half-light radii of each galaxy when searching for substructure using this method.
For regions beyond two half-light radii, we adopted a different approach. We smoothed the stellar distributions using Kernel Density Estimation (KDE), which works by changing the probability density function of each object from a delta function to a pre-defined shape or "kernel" \citep{silverman86}. We divided each field into a 100x100 grid and estimated the surface density of the smoothed data at the center of the individual grid cells. We then generated the number of contaminating objects predicted by the background surface density and spread them randomly across the field.  We smoothed these randomly-positioned background objects with KDE and then subtracted the smoothed background surface density from the smoothed surface density estimate of the actual data to obtain residuals at each location. 

To determine the significance of any substructures in these outlying areas, we compared the residuals to the standard deviation of the smoothed background surface density made from randomly generated objects and 
distributed 
across the field. We avoided classifying the stellar system itself as an over-density by using the same prescription adopted in other studies \citep[e.g.,][]{roderick15a}, where any over-densities greater than 4-$\sigma$ are considered part of the stellar system and are excluded from further analysis. 
Applying this alternative method allows significant over-densities that are well-separated from the main body of the galaxy to be readily identified as a possible fingerprint of substructure.

\subsection{Results}
\label{sec: substructure results}

\subsubsection{Substructure in the RGB Population of Lac~I}
\label{sec: substructure laci}  

As detailed in Section~\ref{sec: substructure methods}, we generated residual maps based on comparing the stellar surface densities within the inner two half-light radii of each galaxy to the surface densities predicted by the best-fitting structural models.  Figure~\ref{fig:substructure.laci.inner} shows the residual maps for the inner regions of Lac~I constructed using the exponential model and the Sersic model.  In the residual map for the exponential model, a string of 2.5- to 3-$\sigma$ over-densities appears, running upward just west of the image center.  The center of the galaxy also shows an under-dense region, which is likely a result of the crowding and foreground star removal mentioned earlier. This under-dense region is less present in the residual map for the Sersic fit, likely because the lower Sersic index provides a better fit to the data there.  The chain of overdensities along the west side of the galaxy still appears, at significance levels of $\sim$2--3-$\sigma$.  Less significant under-dense regions to the north and southeast of the center have expanded in size in this map, but lack the same cohesion of the chain on the western side, suggesting they may be fluctuations rather than genuine substructure.  The overall conclusion is that Lac~I shows an excess of RGB stars in the area between one and two 
times the half-light radius that is not accounted for by the best-fitting models. That these over-densities 
are connected by less significant over-dense regions, suggests that the western chain could be a coherent substructure.

Figure~\ref{fig:substructure.laci.outer} shows the residuals map for the outer regions of Lac~I, beyond two half-light radii from the galaxy center. There are areas with densities more than 2-$\sigma$ above the background around the main body of the galaxy, as well as fluctuations at the $\sim$1-$\sigma$ level in regions beyond the galaxy.  The only notable structure is a 1-2-$\sigma$ filament towards the southeast of the galaxy. This filament is part of the 1-$\sigma$ contour that encompasses the galaxy itself, which suggests that it could be an extended portion of the Lac~I RGB star population rather than merely a fluctuation in the background. Although the significance of the filament is low, further study of the stars in this region may be warranted. 

\begin{figure*}[h]
\plottwo{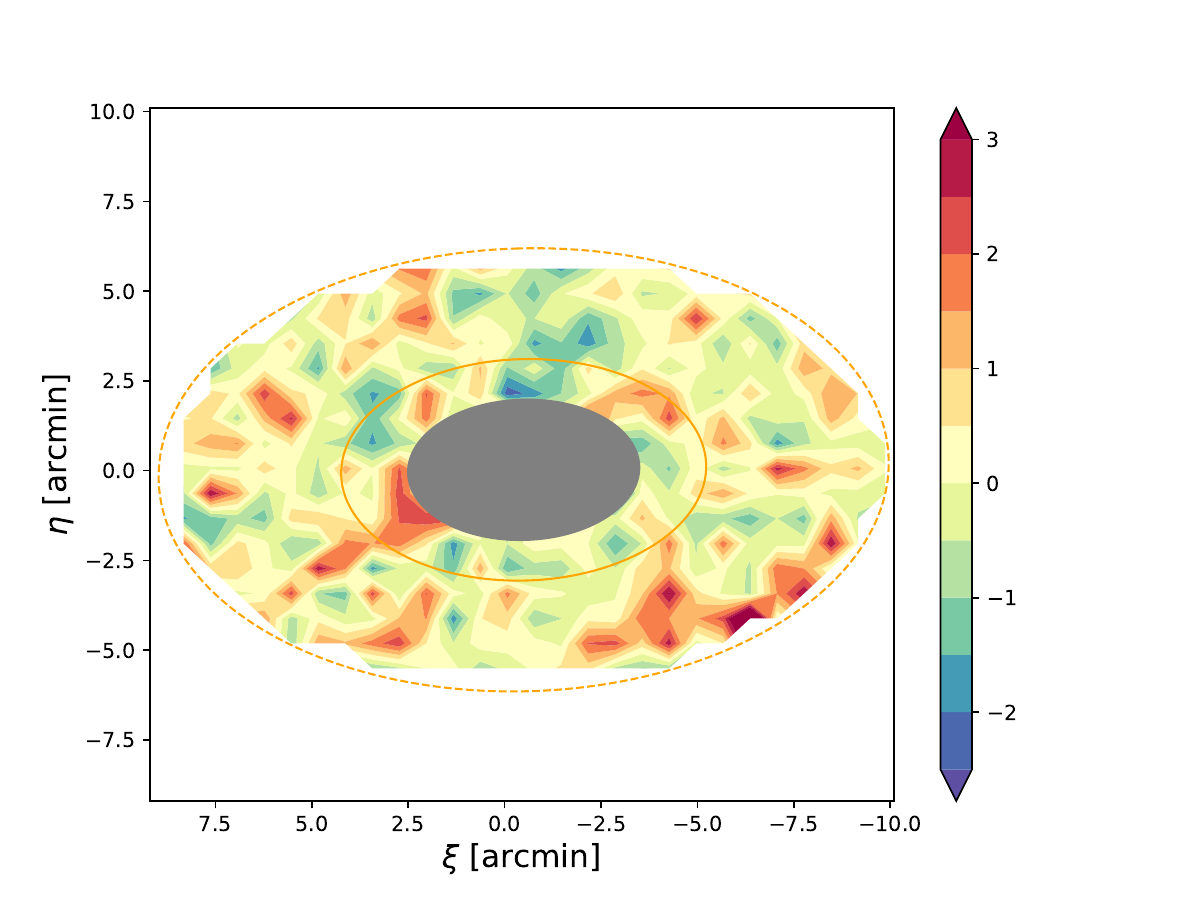}{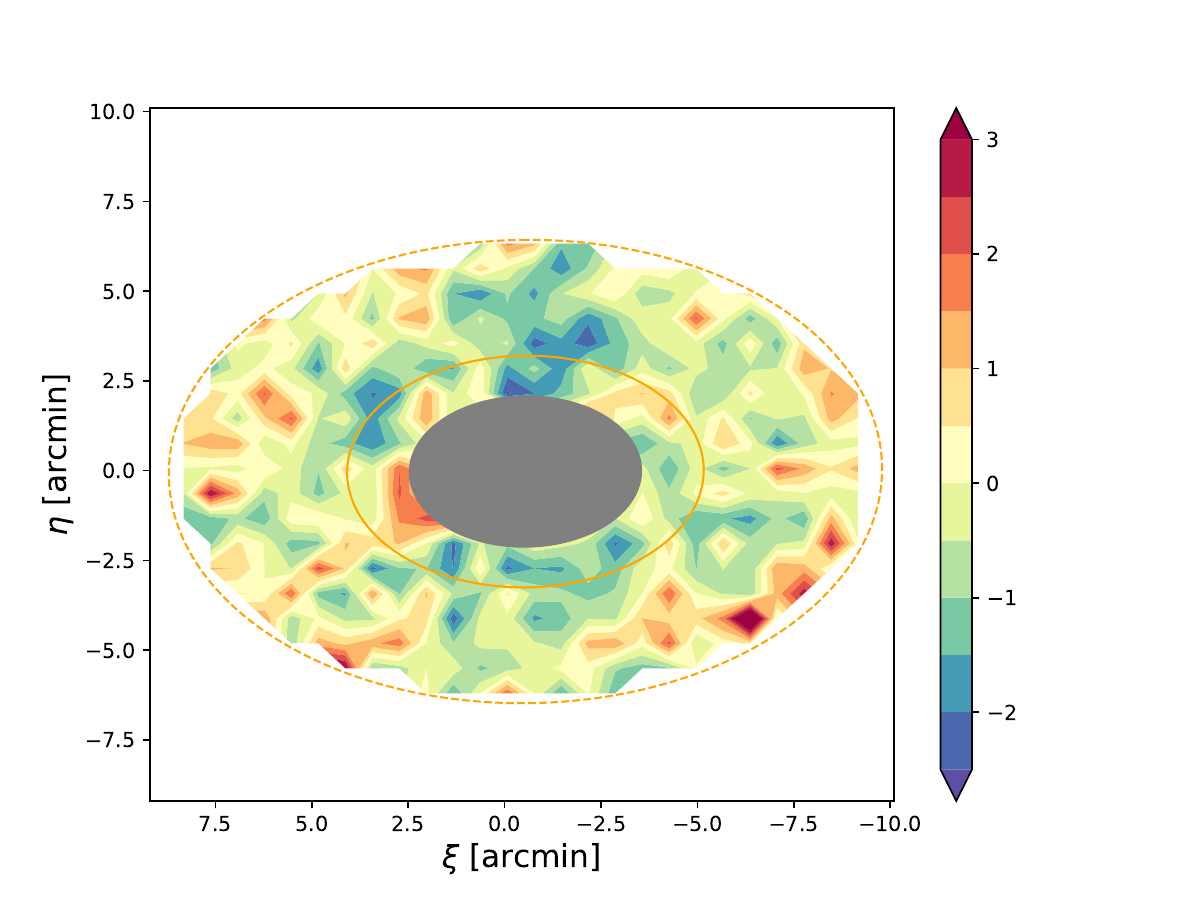}
\caption{The left panel shows the residuals for the regions of Cas~III inside two half-light radii, calculated from the exponential model (left) and the Sersic model 
(right), produced and plotted in the same way as described in the caption for Figure~\ref{fig:substructure.laci.inner}.
The solid and dashed ellipses mark one and two half-light radii, respectively. 
The central regions of Cas~III were affected by stellar crowding, so were excluded from the substructure search process (see Sec.~\ref{sec: substructure casiii}; these areas are shown gray. North is up and east is left.}
\label{fig:substructure.casiii.inner}
\end{figure*}

\begin{figure*}[h]
\plotone{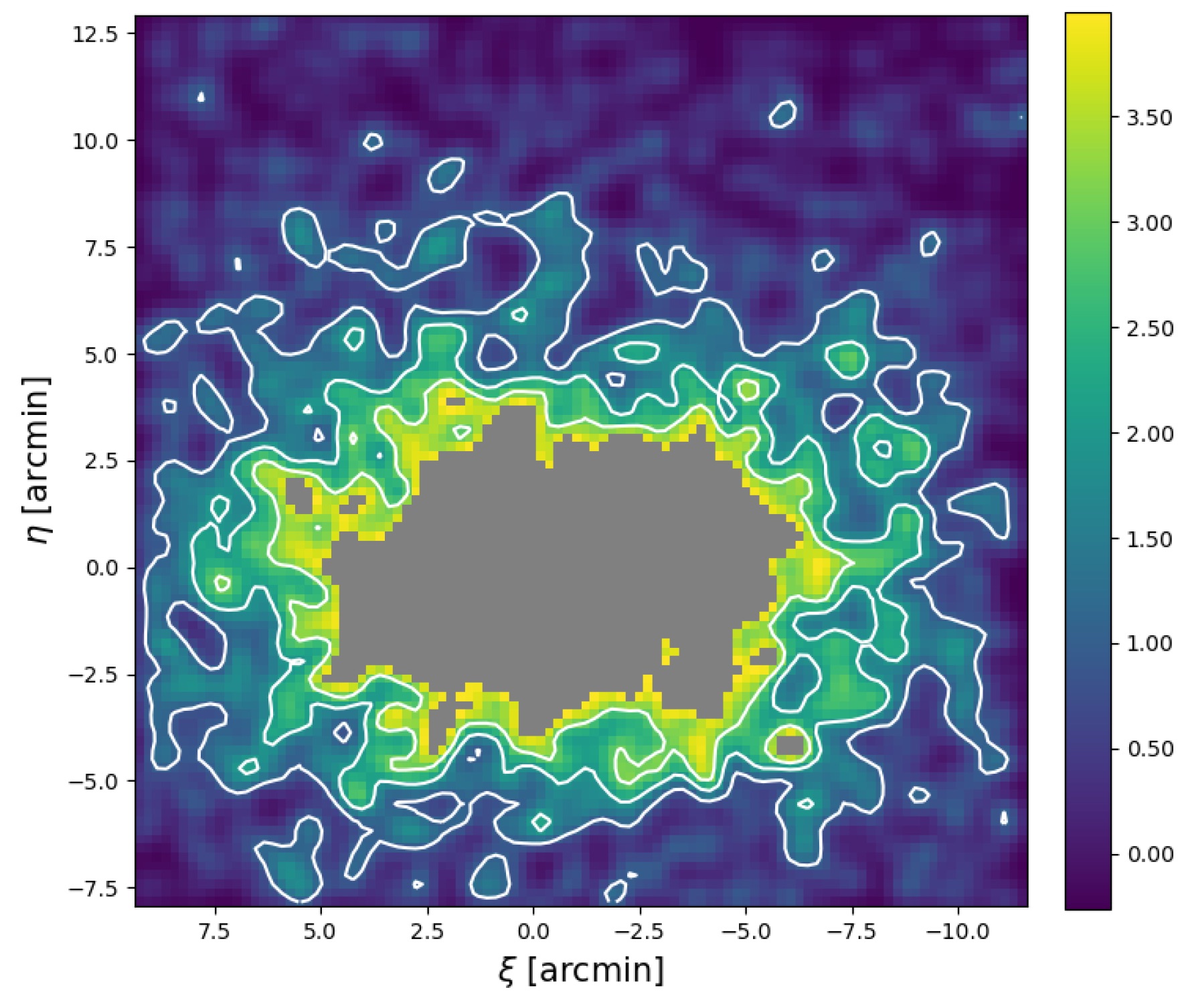}
\caption{
The residual map in the region beyond two half-light radii for Cas~III.  The map was generated as described in Section~\ref{sec: substructure methods}. The main body of the galaxy, where the surface densities are above 4-$\sigma$, is masked in gray. The white contours encompass 1-, 2-, and 3-$\sigma$ areas. 
North is up and east is to the left.}
\label{fig:substructure.casiii.outer}
\end{figure*}

\subsubsection{Substructure in the RGB Population of Cas~III}
\label{sec: substructure casiii}

The structural parameter fitting for Cas~III was impacted by both stellar crowding 
and by the inadequate spatial coverage of our images in the outer parts of the galaxy. To address the stellar crowding issue,
we decided to mask out the inner $\sim$3$\arcmin$ of the galaxy and leave it out of the substructure search completely.  Although we did find that our derived structural parameters for Cas~III disagreed with previously-measured values because of the limited field-of-view of our images, the parameters we derived nevertheless provide us with an accurate description of the observed stellar distribution of Cas~III as it appears in our data set. Therefore we can still examine the deviations from our best-fitting models to look for evidence of substructure in the areas of the galaxy that are beyond the inner, masked region.

The residuals maps for the inner regions of Cas~III, created with the best-fitting exponential and Sersic models, are displayed side-by-side in Figure~\ref{fig:substructure.casiii.inner}.  The area with the worst stellar crowding is shown 
in gray. 
Two over-dense regions appear just outside this masked area, in the northwest and southeast directions.  It is easy to imagine that these two regions could connect across the masked portion of the galaxy, but without further information we cannot say whether this is the case or simply a trick of the eye.
An additional structure with modest ($\sim$2$-$3-$\sigma$) significance extends along the southwest side of the ellipse that marks two half-light radii.  

The substructure map for the outer regions of Cas~III is shown in Figure~\ref{fig:substructure.casiii.outer}. For the most part the isodensity contours match the general shape of the galaxy, showing no obvious tidal features in the background regions.  However, to the southwest of the galaxy there appears a $\sim$4-$\sigma$ feature encompassed by the 2-$\sigma$ contour but otherwise isolated from the rest of the regions with surface densities greater than 3-$\sigma$.  This is likely the same over-density that appears
along the southwest side of the galaxy inside the ellipse that marks two 
half-light radii (Fig.~\ref{fig:substructure.casiii.inner}).  This is not an area of the image that was otherwise masked or affected by stellar crowding, and the method to identify substructure in the outer regions of the images does not depend on the galaxy model-fitting results,
so this makes the case for this feature being a genuine substructure more convincing.

\begin{figure*}[h]
\plottwo{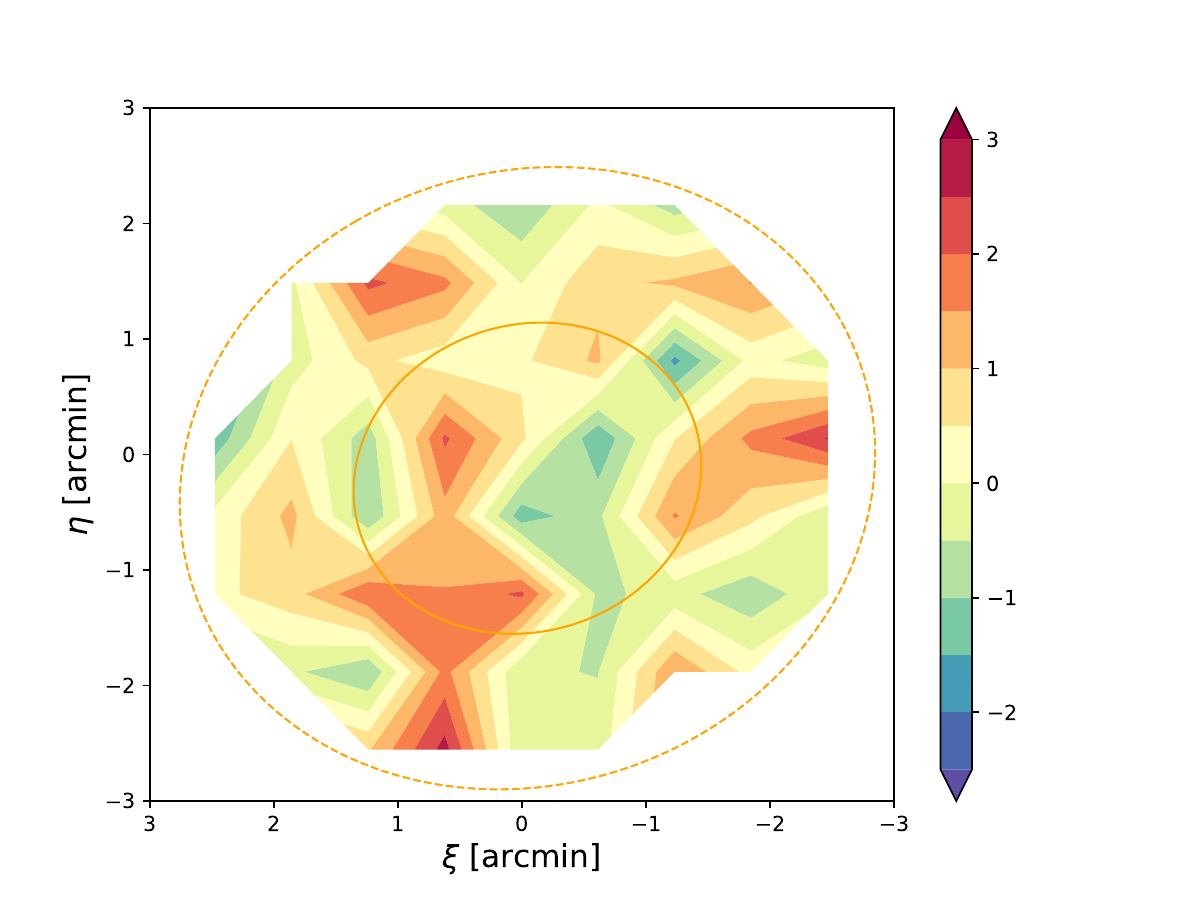}{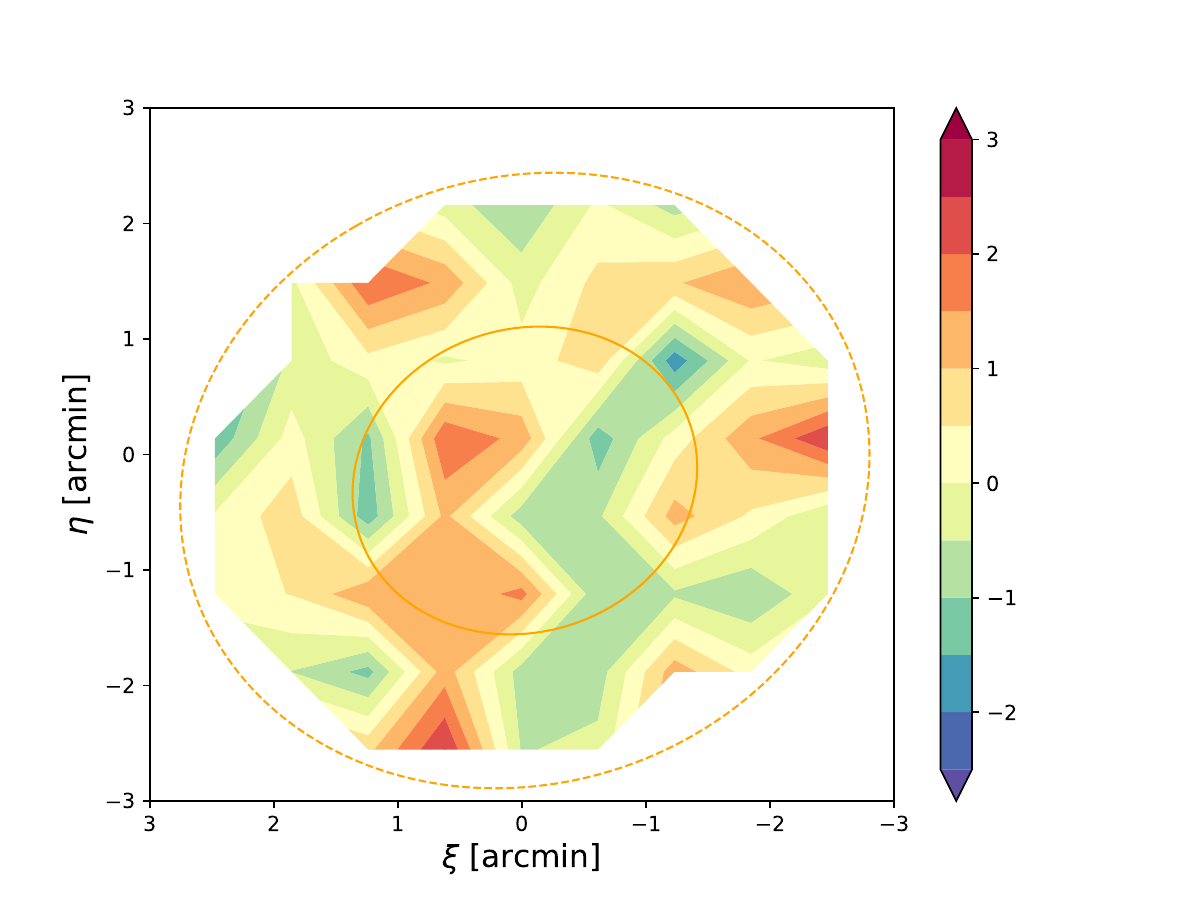}
\caption{
Residuals maps for the portion of Per~I inside two half-light radii, constructed using the exponential model (left) and the Sersic model (right), produced and plotted in the same way as described in the caption for Figure~\ref{fig:substructure.laci.inner}. The solid and dashed ellipses mark one and two half-light radii, respectively. North is up and east is left.}
\label{fig:substructure.peri.inner}
\end{figure*}

\begin{figure*}[h]
\plotone{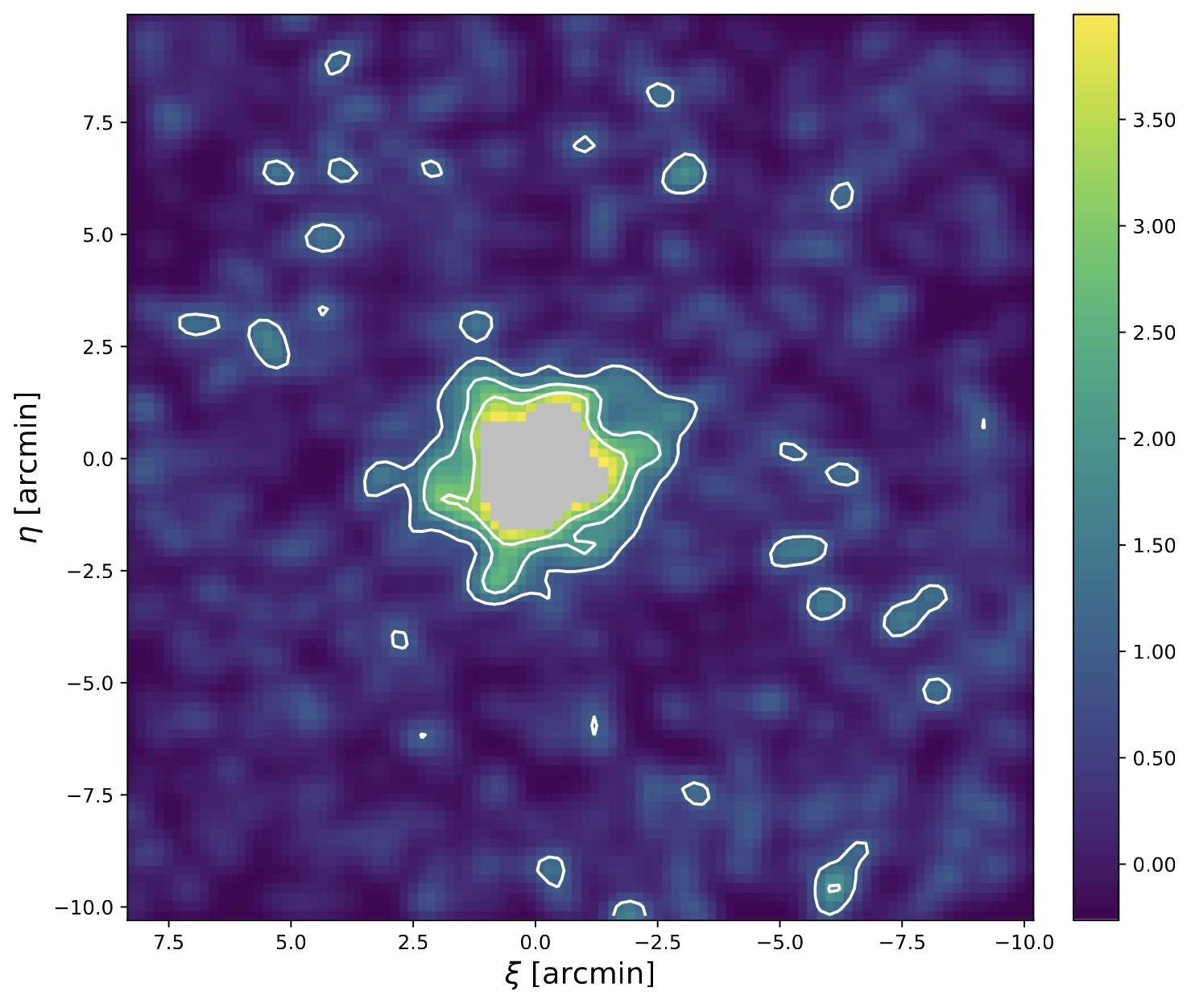}
\caption{
The residuals map for the outer regions in the Per~I field.  The 1-, 2-, and 3-$\sigma$ contours are marked with white lines. The map was generated as described in Section~\ref{sec: substructure methods}. The areas with surface densities greater than 4-$\sigma$ above the background surface density are masked in gray, as they are considered part of the galaxy. 
North is up and east is to the left.}
\label{fig:substructure.peri.outer}
\end{figure*}

\subsubsection{Substructure in the RGB Population of Per~I}
\label{sec: substructure peri}  

Figure~\ref{fig:substructure.peri.inner} shows the residuals maps for the inner regions of Per~I, for both the exponential and Sersic profiles.  The strongest over-density occurs south of the galaxy near the ellipse that marks two half-light radii and reaches a significance of $\sim$2$-$3.5-$\sigma$ in the map made with the exponential model, and a slightly lower significance in the Sersic model map.  There are no other coherent patterns in the residuals or notable features with significance above 2-$\sigma$. Overall the maps seem to reflect a fairly pristine dwarf galaxy.  Per~I lies $\sim$26 degrees ($\sim$350~kpc) from M31 and it seems plausible that it has not had sufficient interaction with its parent galaxy or M31's other satellites to generate the type of tidal features that our methods would reveal. 

Most of the features in the residuals map for the outer regions of Figure~\ref{fig:substructure.peri.outer} appear to be random fluctuations, with no obvious coherent substructures present. However, along the outer edges of the main body of the galaxy there exists a very slight extended shelf along the northwest side as well as a small, 2-$\sigma$ filament extending south away from the galaxy.  While these small features may hint at the presence of modest substructure, additional follow-up observations would be required to confirm that they are genuine evidence of interactions or some other disturbance in the galaxy.

\section{Summary and Conclusions}
\label{sec: summary}

Studies that quantify the properties of the dwarf satellites of the
Milky Way and M31 provide useful insight into the structure and
evolution of dwarf galaxy populations in group environments. In a
companion work to our imaging study of the M31 dwarf satellite galaxy
Lac~I \citep{rhode17}, we have presented measurements of the
photometric and structural properties of the galaxies Cas~III and
Per~I.  All three of the dwarf satellite galaxies we imaged were
discovered in Pan-STARRS data \citep{martin13a,martin13b} and all
three have total magnitudes in the range $M_V$$\sim$$-$9 to $-$12 mag,
faint central surface brightnesses, and sky positions more than 10
degrees away ($\sim$150$-$350~kpc) from their parent galaxy in M31's
outer halo.

We observed the galaxies with the pODI camera on the WIYN 3.5-m telescope and produced CMDs that reach roughly 2$-$3 magnitudes fainter than the photometry in the discovery papers, although our CMDs are much shallower than those derived from HST photometry by \citet{martin17a} and \citet{weisz19a}, which covers the inner portions of the galaxies and reaches below the galaxies' HBs.
Our CMDs each show a prominent RGB sequence and we use them to derive distances to the galaxies using the TRGB method.  Our estimated distance modulus values for Cas~III and Per~I are 24.62$\pm$0.12 mag and 24.47$\pm$0.13 mag, respectively. These values are consistent within the errors with the TRGB distances derived from the HST photometry in \citet{weisz19a}.

We traced the RGB star populations out to approximately 2.5$-$3 half-light radii from the galaxy centers and used an MLE technique to fit both exponential and Sersic function models to the data and estimate the ellipticity, position angle, half-light radius, Sersic index, and central coordinates of each galaxy.  We also derived the total magnitude and central surface brightness and calculated the $M_{\rm HI}$/$L_V$ ratio for each galaxy, confirming that both galaxies are gas-poor.  Our results for the structural parameters of Per~I agree with those from previous work, but with substantially smaller errors on some parameters.  Our structural parameters for Cas~III are different than those derived in the \citet{martin13a} discovery paper. We explore the reasons for the discrepancy and find that it is likely caused by the insufficient spatial coverage of our WIYN data, which falls just short of reaching to three times the half-light radius of the galaxy.

We use our data to revisit a result from Martin et al. (2014), who examined where these galaxies lie in both the mass-size and L-Z planes, compared to expectations for dark-matter-dominated dwarf galaxies and to the 
observed properties of the 
dwarf satellite galaxy populations in the Local Group.  Because our values of the total magnitude and half-light radius are similar to theirs (i.e., within $\sim$1$-$2 times the uncertainty in their measurements), our results do not change their conclusions that Lac~I, Cas~III and Per~I have properties that are comparable to those of other dwarf spheroidal satellite galaxies in the Local Group.

Finally, we carried out a careful analysis to search for substructure within the RGB star populations of all three M31 satellite galaxies in our WIYN data set:  Lac~I, Cas~III, and Per~I.  We developed and used different approaches to find substructures in the inner parts versus the outer parts of the galaxies. 
Our analysis of the Lac~I RGB star distribution
revealed a chain of over-densities along the west side of the galaxy that reached 2.5$-$3-$\sigma$ above the expected surface density.  The over-densities are connected by lower-significance regions that together may comprise a single coherent substructure that spans most of the extent of the galaxy.  In the outer parts of Lac~I, a low-significance filament stretches in the direction of M31 but no other structures are visible.  Cas~III has two over-densities that extend from the center toward the southeast and northwest sides of the galaxy; stellar crowding within the central 3$\arcmin$ of the galaxy makes it impossible to say whether these structures are isolated or a single coherent feature. A modestly over-dense region appears to the southwest of Cas~III, at a radial distance of about twice the half-light radius; this feature was found in the results from both of the approaches we used for detecting substructure. The RGB star population of Per~I shows little evidence of substructure either within two half-light radii or in the surrounding field.  Located $\sim$350~kpc east of M31, this galaxy is the most isolated of the three dwarf satellites in our sample and may simply not have had the tidal interactions necessary to form substructure within its stellar population.  

Detailed studies of dwarf galaxy satellites that include deep, wide-field photometric measurements and investigations of substructure in resolved stellar populations 
will be facilitated by upcoming imaging surveys and missions like the Rubin Observatory Legacy Survey of Space and Time \citep{ivezic19} and the Nancy Grace Roman space telescope \citep{spergel15,akeson19a}. Such studies should prove useful for further developing our understanding of how 
dwarf galaxies in a range of group environments
interact with, and are affected by, their more massive parents.

\begin{acknowledgments}
The authors wish to thank the anonymous referee for providing valuable comments and suggestions for improvements to the manuscript.  We are grateful to the staff of the WIYN 3.5-m Observatory and Kitt Peak National Observatory for their help during our WIYN pODI observing runs. We also thank the staff members at WIYN, NOIRLab, and Indiana University Pervasive Technology Institute for designing and implementing the ODI Pipeline, Portal, and Archive (ODI-PPA) and assisting us with the pODI data reduction.  We made use of the odi-tools python package, written by William Janesh and Owen Boberg, to produce final stacked science-ready WIYN pODI images for this study. K.L.R. and N.J.S. were supported by NSF Astronomy \& Astrophysics Research Grant number AST-1615483 during part of the period when this research was carried out.  Research by D.C. is supported by NSF grant AST-1814208. K.S. acknowledges support from the Natural Sciences and Engineering Research Council of Canada (NSERC). We are grateful to the Indiana University (IU) College of Arts and Sciences for funding IU's share of the WIYN 3.5-m telescope. This research has made use of the NASA/IPAC Extragalactic Database (NED), which is funded by the National Aeronautics and Space Administration and operated by the California Institute of Technology.
\end{acknowledgments}

\software{ODI-PPA \citep{gopu14}, QuickReduce pipeline \citep{kotulla14}, DAOPHOT \citep{stetson87}, ALLFRAME \citep{stetson94}, emcee \citep{foreman-mackey13a}}

\bibliography{mybib}{}
\bibliographystyle{aasjournal}

\clearpage
\begin{deluxetable}{lccccrccc}
\tabletypesize{\footnotesize}
\tablecaption{Catalog of Resolved Stars in the WIYN Images of Cas~III \label{tab:phot.casiii}}
\tablehead{
\colhead{\#}& \colhead{$\alpha$ (2000)}& \colhead{$\delta$ (2000)}&
\colhead{$g_o$}& \colhead{$\sigma_{g}$}& \colhead{$A_g$}
& \colhead{$i_o$}&\colhead{$\sigma_{i}$} & \colhead{$A_i$}\\
\colhead{}& \colhead{(deg)}& \colhead{(deg)}&
\colhead{(mag)}& \colhead{(mag)}& \colhead{(mag)}
&\colhead{(mag)}&\colhead{(mag)}&\colhead{(mag)} }
\startdata
1 & 8.68568 & +51.744674 & 21.09 & 0.01 & 0.62 & 19.92 & 0.01 & 0.32\\
2 & 8.68572 & +51.752808 & 22.75 & 0.01 & 0.61 & 19.95 & 0.00 & 0.31\\
3 & 8.68608 & +51.735231 & 24.13 & 0.05 & 0.62 & 23.00 & 0.07 & 0.32\\
4 & 8.68633 & +51.757495 & 23.76 & 0.04 & 0.61 & 22.48 & 0.05 & 0.31\\
5 & 8.68665 & +51.753527 & 24.18 & 0.06 & 0.61 & 23.00 & 0.06 & 0.31\\
6 & 8.68674 & +51.731567 & 22.37 & 0.01 & 0.62 & 22.06 & 0.02 & 0.32\\
7 & 8.68720 & +51.752180 & 21.67 & 0.01 & 0.61 & 21.12 & 0.01 & 0.32\\
8 & 8.68778 & +51.731062 & 25.59 & 0.17 & 0.62 & 24.44 & 0.21 & 0.32\\
9 & 8.68788 & +51.700293 & 26.20 & 0.31 & 0.65 & 24.57 & 0.25 & 0.34\\
10 & 8.68789 & +51.734888 & 23.98 & 0.04 & 0.62 & 22.61 & 0.04 & 0.32\\

\enddata
\tablecomments{This table is available in its entirety in a machine-readable form in the online journal. A small portion of the data is shown as an example of the form and content of the table.}
\end{deluxetable}

\begin{deluxetable}{lccccrccc}
\tabletypesize{\footnotesize}
\tablecaption{Catalog of Resolved Stars in the WIYN Images of Per~I \label{tab:phot.peri}}
\tablehead{
\colhead{\#}& \colhead{$\alpha$ (2000)}& \colhead{$\delta$ (2000)}&
\colhead{$g_o$}& \colhead{$\sigma_{g}$}& \colhead{$A_g$}
& \colhead{$i_o$}&\colhead{$\sigma_{i}$} & \colhead{$A_i$}\\
\colhead{}& \colhead{(deg)}& \colhead{(deg)}&
\colhead{(mag)}& \colhead{(mag)}& \colhead{(mag)}
&\colhead{(mag)}&\colhead{(mag)}&\colhead{(mag)} }
\startdata
1 & 45.12081 & +41.141294 & 25.40 & 0.27 & 0.36 & 24.89 & 0.38 & 0.19\\
2 & 45.12084 & +41.143717 & 26.35 & 0.32 & 0.36 & 25.31 & 0.30 & 0.19\\
3 & 45.12124 & +41.132735 & 24.96 & 0.09 & 0.36 & 24.55 & 0.14 & 0.19\\
4 & 45.12144 & +41.128517 & 23.38 & 0.02 & 0.37 & 21.51 & 0.01 & 0.19\\
5 & 45.12161 & +41.146342 & 25.23 & 0.10 & 0.36 & 24.40 & 0.11 & 0.19\\
6 & 45.12172 & +41.143182 & 26.08 & 0.22 & 0.36 & 25.14 & 0.24 & 0.19\\
7 & 45.12186 & +41.115449 & 26.42 & 0.31 & 0.37 & 25.04 & 0.22 & 0.19\\
8 & 45.12249 & +41.093886 & 25.17 & 0.10 & 0.38 & 24.68 & 0.14 & 0.19\\
9 & 45.12262 & +41.116702 & 25.47 & 0.14 & 0.37 & 25.33 & 0.29 & 0.19\\
10 & 45.12283 & +41.104684 & 23.24 & 0.02 & 0.37 & 20.24 & 0.01 & 0.19\\

\enddata
\tablecomments{This table is available in its entirety in a machine-readable form in the online journal. A small portion of the data is shown as an example of the form and content of the table.}
\end{deluxetable}

\begin{deluxetable}{lrrr}
\tabletypesize{\footnotesize}
\tablecolumns{2}
\tablewidth{0pt}
\tablecaption{Properties of Lac~I, Cas~III, and Per~I}
\tablehead{
\colhead{Property}& \colhead{Lac~I} & \colhead{Cas~III} & \colhead{Per~I}  
}
\startdata
Right Ascension ($\alpha$) & 22h 58m 13.3s  & 00h 35m 59.4s & 03h 01m 23.6s \\
Declination ($\delta$)  & $+$41$^{\circ}$ 17$'$ 53.5$''$ & $+$51$^{\circ}$ 33$'$ 35$''$ & 40$^{\circ}$ 59$'$ 18$''$ \\
$M_V$ (mag)	 	& $-$11.4 $\pm$ 0.3 & $-$12.5 $\pm$ 0.2 & $-$9.1 $\pm$ 0.3 \\
$M_g$ (mag) 	& $-$11.3 $\pm$ 0.2	& $-$12.1 $\pm$ 0.1 & $-$8.8 $\pm$ 0.3 \\
($g$$-$$i$)$_0$ (mag) & 0.7 $\pm$ 0.3 & 1.4 $\pm$ 0.3 & 1.0 $\pm$ 0.3 \\
$(m-M)_0$ (mag)  & 24.44 $\pm$ 0.11 & 24.62 $\pm$ 0.12 & 24.47 $\pm$ 0.13 \\
$D$ (kpc) & 773 $\pm$ 40 & $839^{+48}_{-45}$ & $783^{+48}_{-45}$  \\
$D_{\rm M31}$ (kpc) & 264 $\pm$ 6 &  $156^{+16}_{-13}$ & $351^{+17}_{-16}$ \\
$\mu_{V,0}$ (mag/arcsec$^2$) & 24.8 $\pm$ 0.3 &  24.9 $\pm$ 0.3 & 25.7 $\pm$ 0.3\\ 
$r_h$ (arcmin) &  3.13 $\pm$ 0.05 & 4.73 $\pm$ 0.09 &  1.40$^{+0.07}_{-0.06}$ \\
$r_h$ (pc) &  704 $\pm$ 11 & 1154 $\pm$ 22 &  319 $\pm$ 16 \\
Ellipticity ($\epsilon$)& 0.40 $\pm$ 0.01 &  0.35 $\pm$ 0.01 & 0.09 $\pm$ 0.06\\
Position angle ($\theta$, $^{\circ}$ East of North) & $-$63 $\pm$ 1 & 91 $\pm$ 2 & $-$59$^{+23}_{-20}$ \\
$M_{HI}/L_V$ ($M_\sun$/$L_\sun$) & $<$0.06 & $<$0.009 & $<$0.22\\
${\rm [Fe/H]}$ (dex) & $-$2.0 $\pm$ 0.1 & $-$1.7 $\pm$ 0.1 & $-$2.0$\pm$0.2 \\
\enddata
\tablecomments{RA and Dec coordinates are from the NASA/IPAC Extragalactic Database (NED). The absolute magnitudes are calculated using our TRGB distance measurements. The structural parameters listed here are for the exponential model ($n=$1); Table~\ref{table:structural params all} shows the full set of structural parameters for both the exponential and Sersic ($n=$free) models. The $r_h$ values given in parsecs are calculated using $r_h$ from our exponential model fit and our derived TRGB distance.  For convenience we list here several properties of Lac~I that were derived in the \citet{rhode17} study (namely, position, total absolute magnitude, color, distance modulus, distance, central surface brightness, and $M_{HI}/L_V$); the structural parameters ($r_h$, ellipticity, position angle) for Lac~I shown in this table are from the current study, but in any case are consistent with those derived by \citet{rhode17}. We also list} the [Fe/H] estimates for each galaxy from the \citet{martin14} spectroscopic study.
\label{tab:properties all}
\end{deluxetable}

\begin{deluxetable}{lcccc}
\tabletypesize{\footnotesize}
\tablecaption{Distance Modulus Estimates for Lac~I, Cas~III, and Per~I \label{tab:distance comparison}}
\tablehead{
\colhead{Galaxy}& \colhead{TRGB (HST)}& \colhead{HB (HST)}&
\colhead{RRL (HST)}& \colhead{TRGB (WIYN)}\\
\colhead{}& \colhead{(mag)}& \colhead{(mag)}&
\colhead{(mag)}& \colhead{(mag)} }
\startdata
Lac~I & 24.51$^{+0.03}_{-0.02}$ & 24.50$^{+0.05}_{-0.04}$ & 24.36$\pm$0.05 & 24.44$\pm$0.11\\
Cas~III & 24.57$^{+0.08}_{-0.03}$ & 24.70$\pm$0.04 & 24.52$\pm$0.06 & 24.62$\pm$0.12\\
Per~I & 24.49$^{+0.14}_{-0.28}$ & 24.39$^{+0.05}_{-0.03}$ & 24.24$\pm$0.06 & 24.47$\pm$0.13\\
\enddata
\tablecomments{Distance modulus values for each galaxy are drawn from the TRGB and HB measurements in \citet{weisz19a} (columns 2 and 3, respectively), the RR~Lyrae variable star  measurements in \citet{savino22a} (column 4), and the ground-based WIYN photometry TRGB measurements in \citet{rhode17} and the current paper (column 5).}
\label{tab:distances}
\end{deluxetable}

\begin{table}
\caption{Structural Parameters for Lac~I, Cas~III, and Per~I}
\begin{tabular}{lrrrrrr}
 & \multicolumn{2}{c}{{Lac~I}} & \multicolumn{2}{c}{{Cas~III}} & \multicolumn{2}{c}{{Per~I}} \\ \cline{2-7} 
{Parameter} & \multicolumn{1}{c}{{n free}} & \multicolumn{1}{c}{{n=1}} & \multicolumn{1}{c}{{n free}} & \multicolumn{1}{c}{{n=1}} & \multicolumn{1}{c}{{n free}} & \multicolumn{1}{c}{{n=1}} \\ 
\cline{1-7}
Ellipticity ($\epsilon$) & 0.41$\pm$0.01 & 0.40$\pm$0.01 & 0.30$\pm$0.01 & 0.35$\pm$0.01 & 0.09$\pm$0.06 & 0.09$\pm$0.06 \\
PA ($\theta$) (deg) & $-$64$\pm$1 & $-$63$\pm$1 & 91$\pm$2 & 91$\pm$2 & $-58^{+25}_{-21}$ & $-59^{+23}_{-20}$ \\
$r_h$ (arcmin) & 3.18$\pm$0.05 & 3.13$\pm$0.05 & 4.63$\pm$0.06 & 4.73$\pm$0.09 & 1.39$\pm$0.07 & $1.40^{+0.07}_{-0.06}$ \\
Sersic index ($n$) & 0.83$\pm$0.03 & ... & 0.57$\pm$0.02 & ... & 0.7$\pm$0.1 & ... \\
$\Delta$RA (arcmin) & $-$0.27$\pm$0.04 & $-$0.28$\pm$0.04 & $-$0.54$\pm$0.07 & $-$0.49$\pm$0.06 & $-$0.02$\pm$0.06 & $-$0.04$\pm$0.06 \\
$\Delta$Dec (arcmin) & 0.48$\pm$0.03 & 0.48$\pm$0.03 & $-$0.02$\pm$0.05 & 0.02$\pm$0.05 & $-$0.22$\pm$0.06 & $-$0.21$\pm$0.06 \\
\cline{1-7}
\end{tabular}
\tablecomments{Position angle (PA) values are measured from North through East. The $\Delta$RA and $\Delta$Dec values are relative to the galaxy positions given in 
Table~\ref{tab:properties all}, which are drawn from NED; the NED position for Lac~I is RA = 344.567917$^\circ$, Dec = 41.291111$^\circ$.}
\label{table:structural params all}
\end{table}

\end{document}